\documentclass[acmtog]{acmart}
\usepackage{booktabs} 

\usepackage{listings}
\usepackage[export]{adjustbox}

\usepackage[ruled]{algorithm2e} 

\SetAlFnt{\small}
\SetAlCapFnt{\small}
\SetAlCapNameFnt{\small}
\SetAlCapHSkip{0pt}

\ccsdesc[500]{Computing methodologies~Mesh models}
\ccsdesc[500]{Computing methodologies~Mesh geometry models}
\ccsdesc[300]{Computing methodologies~Shape analysis}

\keywords{volume meshing, numeric robustness, representability}

\usepackage{dsfont}
\usepackage{amsthm}
\usepackage{overpic}
\usepackage{contour}
\contourlength{1pt}
\usepackage{xspace}
\usepackage{sidecap}
\usepackage{rotating}
\usepackage[normalem]{ulem}
\usepackage{booktabs} 
\usepackage{caption}
\usepackage{subcaption}
\usepackage{microtype}
\usepackage{booktabs}
\usepackage{enumitem}
\usepackage{amsthm}
\usepackage{wrapfig}
\usepackage{multirow}
\usepackage{tabularx}
\usepackage{nicefrac}
\usepackage{chngcntr}
\usepackage{apptools}
\usepackage[super]{nth}
\AtAppendix{\counterwithin{lemma}{section}}

\newlength{\indentlaenge}
\newlength{\mylength}
\newlength{\mylengthzwei}
\newcommand{\rev}[1]{{#1}}

\def\cput(#1,#2)#3{\put(#1,#2){\hbox to 0pt{\hss{#3}\hss}}}
\def\lput(#1,#2)#3{\put(#1,#2){\hbox to 0pt{\hss{#3}}}}
\def\rput(#1,#2)#3{\put(#1,#2){\hbox to 0pt{{#3}\hss}}}

\begin{document}

\title{Constrained Delaunay Tetrahedrization: A Robust and Practical Approach}

\author{Lorenzo Diazzi}
\affiliation{%
  \institution{UniMoRe}
  \country{Italy}
  \institution{IMATI - CNR}
  \country{Italy}
  }
\email{lorenzo.diazzi@unimore.it}

\author{Daniele Panozzo}
\affiliation{%
  \institution{New York University}
  \country{USA}
  }
\email{panozzo@nyu.edu}

\author{Amir Vaxman}
\affiliation{%
  \institution{The University of Edinburgh}
  \country{UK}
  \orcid{0000-0001-6998-6689}
  }
\email{avaxman@inf.ed.ac.uk}

\author{Marco Attene}
\affiliation{%
  \institution{IMATI - CNR}
  \country{Italy}
  }
\email{marco.attene@ge.imati.cnr.it}

\begin{teaserfigure}
  \centering
  \includegraphics[width=\textwidth]{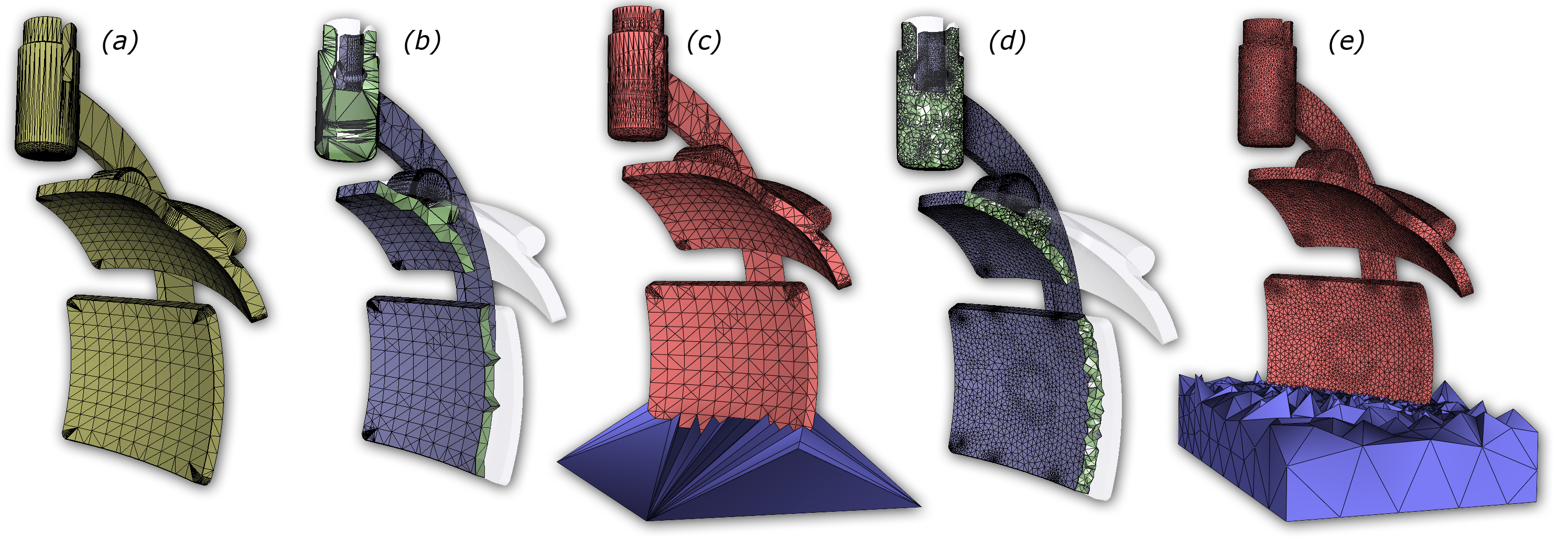}
  \caption{A valid model in the Thingi10k dataset (a) is enriched with Steiner points and tetrahedrized by our method (b, c). The resulting tetrahedrization is then used as a starting point for a standard mesh optimization process (d, e).}
  \label{fig:teaser}
\end{teaserfigure}


\begin{abstract}
We present a numerically robust algorithm for computing the constrained Delaunay tetrahedrization (CDT) of a piecewise-linear complex, which has a 100\% success rate on  the 4408 valid models in the Thingi10k dataset.

We build on the underlying theory of the well-known \texttt{tetgen} software, but use a floating-point implementation based on indirect geometric predicates to implicitly represent Steiner points: this new approach dramatically simplifies the implementation, removing the need for ad-hoc tolerances in geometric operations. Our approach leads to a robust and parameter-free implementation, with an empirically manageable number of added Steiner points. Furthermore, our algorithm addresses a major gap in \texttt{tetgen}'s theory which may lead to algorithmic failure on valid models, even when assuming perfect precision in the calculations.

Our output tetrahedrization conforms with the input geometry without approximations. We can further round our output to floating-point coordinates for downstream applications, which almost always results in valid floating-point meshes unless the input triangulation is very close to being degenerate.



\end{abstract}

\maketitle


\section{Introduction}
\label{sec:introduction}
A Constrained Delaunay Tetrahedrization (CDT) is a compact space subdivision widely used for diverse computer graphics and engineering applications, including the solution of partial differential equations \cite{si_fem},  calculation of shape thickness \cite{CABIDDU2017}, medial axis approximation \cite{dey_ma}, ray tracing \cite{cdt4raytracing}, and many other algorithms requiring an explicit discretization of a volume. Preserving the boundary of the input surface in tetrahedral meshing is often critical, especially for solving PDEs on complex geometric domains, as this allows to adhere to the boundary constraints exactly without loss of accuracy. Boundary-\emph{approximating} tetrahedral meshing algorithms (such as TetWild \rev{\cite{Hu:2018}}, Quartet \cite{isostuff}\rev{\cite{Bridson:2014}}, or CGAL \cite{fabri2009cgal}) need to rely on lossy and potentially not bijective projections to map boundary conditions, which makes them unsuitable for such applications. Our algorithm by definition produces a boundary-preserving CDT.


\paragraph{Constrained Delaunay Triangulations}
Roughly speaking, among all the possible triangulations, the CDT is the one that satisfies the Delaunay criterion \emph{as much as possible} while containing all the input segments (in 2D) and/or facets (in 3D) \cite{alexa_cwdt}.
In 2D, a collection of non-intersecting and non-degenerate segments always admits a valid CDT. This is not the case in 3D, where the existence of a CDT is no longer guaranteed unless we augment the input Piecewise-Linear Complex (PLC) with a set of additional Steiner points \cite{murphy2001point, Shewchuk2002CDT, HSi2005CDT}. Unfortunately,  determining the number and the optimal position of these points is still an open problem \cite{RupSei92}.

\paragraph{Existing Algorithms.}
Provably-correct algorithms (Sect. \ref{sec:related-work}) exist that first augment a PLC with Steiner points, and then calculate a valid CDT. Most of these algorithms are, however, impractical since they introduce a large number of Steiner points. 
\cite{HSi2005CDT} made a major theoretical and practical step forward in computing CDTs by introducing a new algorithm which, in practical cases, introduces a small number of Steiner points. The popular software \texttt{tetgen} is a floating-point implementation of this algorithm.
%
%
To the best of our knowledge, \texttt{tetgen} is the only publicly-available implementation of an algorithm to compute a CDT in 3D; while well-engineered, it fails to produce a CDT on around 8.5\% of the valid models in the Thingi10k dataset. Fig. \ref{fig:teaser} depicts an example \rev{model} that makes \texttt{tetgen} crash, whereas our \rev{method} processes it correctly.

\paragraph{Analyzing failure in \texttt{tetgen}} While it is intuitive to assume that the said failures in \texttt{tetgen} are due to numerical tolerance used in their floating-point implementation, this is not always the case. In Sec. \ref{sec:issues} we introduce and analyze a theoretical issue that results from a tacit and incorrect assumption in the theory of \cite{HSi2005CDT}. We validated this using an exact-number implementation (CORE library \cite{core1999,burnikel1996leda}).


\paragraph{Contribution.}

We introduce a novel algorithm for computing the CDT of a valid PLC that is provably correct and can be implemented based on hardware-accelerated floating-point computation, without sacrificing robustness. Our two main contributions are:

\begin{enumerate}
    \item A novel algorithm, \rev{based on a combination of \cite{HSi2005CDT} and \cite{Shew00GW}}, that avoids the tacit assumption that all the local cavities formed to insert PLC faces can be sufficiently expanded (Sec. \ref{sec:issues}). A key property of our algorithm, which is crucial for the second contribution, is that it does not require using irrational coordinates for Steiner points.
    \item An exact and efficient implementation of our algorithm based on indirect floating-point predicates \cite{MAtteneIndPred}. \rev{We define a new type of implicit point to represent the Steiner vertices (Sec. \ref{sec:fast-implementation}), extend the classical orient3D and inSphere predicates to handle this new point type (Sec. \ref{sec:indirect_preds}), and show that these new predicates are sufficient to implement all the non-trivial checks required by the combined algorithm (App. \ref{app:GWpred}).} 
\end{enumerate}

To the best of our knowledge, our algorithm is the first that successfully computes the CDT for \textbf{all} the 4408 valid PLCs in the Thingi10k dataset. Our implementation combines theoretical correctness, practical robustness, and efficiency thanks to a safe use of floating point arithmetic. 
Our prototype  can process all the said 4408 models in approximately 5 hours on a single CPU core on a standard desktop PC. Our reference implementation is available as an open-source project (Sec. \ref{sec:result_and_discussion}) to foster the adoption of our algorithm and support the many applications which use CDT computation as an internal routine.



\section{Related Work}
\label{sec:related-work}
Subdividing a volumetric domain with tetrahedra has been extensively studied in the last decades. We refer the reader to~\cite{Hu:2018,Hu:2020} for an elaborate literature study, and here focus on the relevant works that mostly target conformity (as much as possible) to valid input boundary surfaces.

\subsection{Delaunay meshing}
\paragraph{Delaunay tetrahedrization}
The Delaunay tetrahedrization (DT) of a point set in space can be calculated using incremental insertion algorithms \cite{bowyer81} \rev{\cite{watson81}} whose output satisfies the so-called \emph{Delaunay criterion}: the interior of the circumsphere of any tetrahedron does not contain any point of the input point set. Incremental insertion algorithms can be implemented both robustly and efficiently \cite{fabri2009cgal}, and modern versions can also be parallelized \cite{marot2019one}.
When the input is a polyhedral surface, one may compute the Delaunay tetrahedrization of its vertices, but the edges and facets are not necessarily represented.

\paragraph{Constrained Delaunay}
Given a polygon in 2D, one can calculate its \emph{constrained} Delaunay triangulation \rev{\cite{lee86}} \cite{chew89}, where both vertices and edges are represented.
Unfortunately obtaining a similar result for 3D polyhedra is substantially more difficult because, as previously mentioned, not all polyhedra admit a CDT, and additional Steiner points may be necessary to obtain a useful result. Some pioneering results in this area were obtained in \cite{geompack}, where the idea was to first subdivide the polyhedron into convex regions whose DT also conforms with the facets. Then individual DTs can be simply merged, but the initial subdivision turned out to be particularly difficult.
\cite{george} and \cite{weatherill} introduced an effective approach which inspired many subsequent works. After having computed the DT of the vertices, this approach first recovers the input \rev{segments} and then reconstructs the facets in a second phase. This was subsequently adopted in \cite{guan}, where an empirically smaller number of Steiner points was used.
\cite{Shewchuk2002CDT} introduced a provably correct algorithm for computing the CDT by \emph{protecting} acute vertices, which amounts to splitting all edges incident to such vertices. Inspired \rev{by} this work, \cite{HSi2005CDT} presented an alternative approach that still protects the vertices, but employs a significantly smaller number of Steiner points to do that. The famous \texttt{tetgen} software implementing this approach \rev{is} the state of the art for the calculation of CDTs in 3D. We compare against it in Section \ref{sec:result_and_discussion}.

\paragraph{Conforming Delaunay} Constrained tetrahedrizations put in as few Steiner points as possible to make the tetrahedrization feasible. However, the elements are not guaranteed to be Delaunay everywhere. Conversely, conforming tetrahedrizations (e.g.,\rev{~\cite{Cohen-Steiner:2002, murphy2001point}}) generate Delaunay tetrahedrizations that conform to a refinement of the boundary to subfaces and subedges, while introducing potentially many Steiner points, and consequently many tetrahedra.
Recently, Alexa \shortcite{alexa_cwdt} observed that in some cases weights can be assigned to the input vertices so that their weighted Delaunay tetrahedrization contains the input simplices with no need of Steiner points. However, such a set of weights may not exist and, when it does, calculating it is impractically slow. 

\paragraph{Delaunay refinement} Given an initial tetrahedrization, one can successively improve it by introducing new vertices at the centers of circumscribing spheres of bad tetrahedra (e.g.,~\cite{Shewchuk1998EThm, Jamin:2015, ruppert:1995}). Such methods are successfully implemented in CGAL Mesh Generation package~\cite{Rineau:2007} and in \texttt{Tetgen}~\cite{shewchuk2014higher} (see Sec.~\ref{sec:delaunay-refinement}). While guaranteeing termination, these methods often admit \emph{slivers}, which are tets with nominally good radius-edge-length ratios, but close to degenerate in terms of volume. Approaches to remove slivers include relaxation~\cite{du:2003, alliez:2005} and perturbation~\cite{tournois:2009}. We note that a recent method~\cite{alexa_harmonic} computes \emph{harmonic} triangulations that minimize the Dirichlet energy, rather than Delaunay, noting that these properties are equivalent in 2D. We compare against CGAL in Section \ref{sec:result_and_discussion}.

\subsection{Non-Delaunay Tetrahedral meshing}

\paragraph{Grid based} A class of methods establishes a uniform grid, or an adaptive octree around an object, which is simple to tetrahedrize. To conform to a boundary surface that is not grid aligned, some methods cut the existing grid cells and tetrahedrize the intersection\rev{~\cite{Bronson:2013, isostuff, Doran:2013, Bridson:2014}}, whereas some deform the grid to match the original boundary~\cite{Molino:2003}. Some guarantees on average element quality exist for low-curvature objects that are sufficiently convex (that is, have a high volume-to-surface ratio). However, element quality either degrades considerably near the surface, or many elements are required. We show a comparison against the representative method Quartet \rev{\cite{Bridson:2014}} in Section \ref{sec:result_and_discussion}.

\paragraph{Advancing front} Some methods (e.g.,\cite{Alauzet:2014, Cuilliere:2013,Haimes:2015,Frey1996DelaunayTU}) start from a given front (the boundary surface) and propagate meshes inwards. While seemingly achieving good element quality near the boundary, these methods suffer from problematic cases when fronts meet in the interior of the volume, making the generation of high-quality elements in these places challenging. We are not aware of any robust and publicly available representative implementation for this category.

\paragraph{In the wild} A considerable portion of methods do not inherently assume the boundary is valid or watertight. Thus, a resulting tet mesh would not necessarily be conforming. One such approach is envelope meshing (e.g.,~\cite{Mandad:2015,Shen:2004,Hu:2018}), where the surface is approximated to some volumetric tolerance, or an ``envelope'' around the original boundary components. 
While these methods solve an inherently different problem, we show a comparison against the representative method TetWild in Section \ref{sec:result_and_discussion}, where we show that the boundary is not exactly represented.

\subsection{Robust geometric predicates}
Numerical robustness is a common issue in many meshing algorithm implementations. If standard floating point arithmetic is used with no specific care, an implementation may easily crash or fail to converge \cite{li2005}.
Implementations can be made robust by using exact arithmetic kernels \cite{fabri2009cgal}. However, this solution is often too slow for practical applications. Alternatively, robustness can be achieved by simply guaranteeing that the program flow is correct while accepting a rounded output \cite{li2005}. The program flow is determined by its branches that, in turn, are governed by the value of geometric predicates analyzing the relative position of points:
when their coordinates are read from an input file, floating-point filtering techniques and adaptive precision can combine speed and robustness \cite{shew97_pred}. This is not sufficient when some of the points being analyzed are constructed by the algorithm. Indeed, in this case, coordinates can be rounded and predicates are no longer guaranteed to give the correct answer. However, if these intermediate points can be expressed as simple combinations of input points, we can still exploit the concept of indirect predicates \cite{MAtteneIndPred} to leverage the efficiency of the floating point hardware. This solution was successfully used to create polyhedral meshes in \cite{DiazziAttene2021}, though the arbitrarily bad shape of cells severely limits their potential application.


\section{Background}
\label{sec:background}

Unless explicitly stated otherwise, all the concepts described in this section are defined in three-dimensional space.

\subsection{Problem statement}
\begin{figure}
    \centering
    \includegraphics[width=\columnwidth]{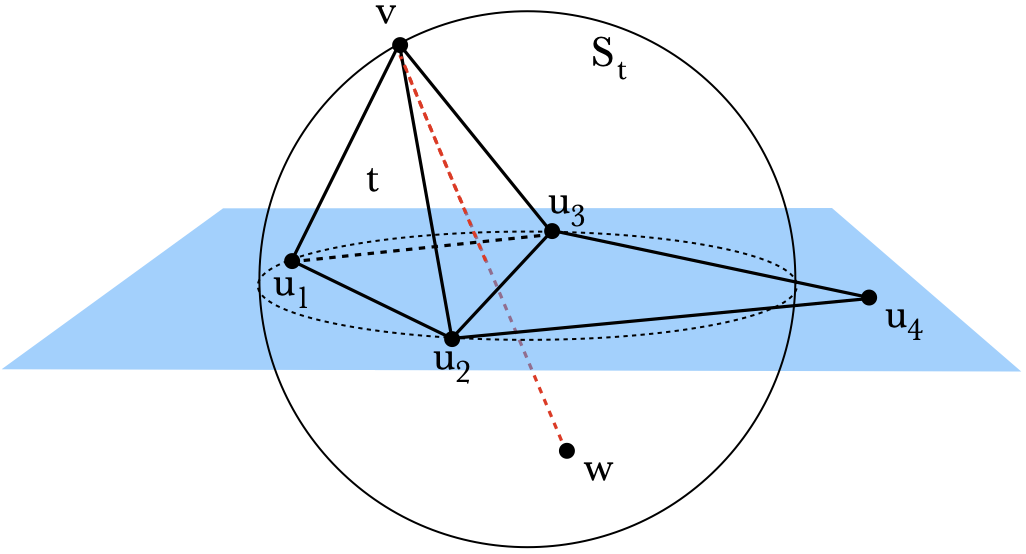}
    \caption{A constrained Delaunay Tetrahedron $t$. The blue region is the supporting plane to  PLC faces $\langle u_1,u_2,u_3 \rangle $ and $\langle u_2,u_3,u_4\rangle $. Vertices $v$ and $w$ belong to opposite half-spaces defined by the plane, and thus $w$ is not visible from the interior of tetrahedron $t=\langle u_1,u_2,u_3,v\rangle $. In contrast, vertex $u_4$ is visible from $t$. $S_t$ is the circumsphere of tetrahedron $t$ that encloses $w$ but not $u_4$. Therefore, $t$ is constrained Delaunay.}
    \label{fig:ConstrDelTet}
\end{figure}

Our input is a piecewise-linear complex (PLC) \cite{miller1996control}. A PLC is a collection of vertices, \rev{segments}, and polygonal facets with the typical characteristics of a complex: the boundary of any element is made of lower-dimensional elements of the PLC, and the set is closed under intersection; that is, the intersection of any two elements of a PLC is either empty or it is the union of other elements of the PLC.
Our objective is to compute a CDT that preserves the original PLC. Specifically, if $P$ is our PLC, we want to compute a tetrahedrization $T$ of the convex hull of $P$ such that $T$ has all, and only, the vertices of $P$, and each facet of $P$ is the union of some triangular facets of $T$. Furthermore, we require that $T$ is constrained Delaunay with regards to  $P$; that means that the interior of the circumsphere of each tetrahedron does not contain any \emph{visible} vertex of $P$ (see Fig. \ref{fig:ConstrDelTet}). In turn, a vertex $v$ is visible from within a tetrahedron $t$ if 
any
point in the interior of  $t$ can be connected to $v$ using a straight segment that does not intersect $P$.

\subsection{Characteristics of a PLC}
As mentioned in the introduction, not all PLCs admit a CDT. Sch\"onhardt's polyhedron \cite{SchonPol} and Chazelle's polyhedron \cite{chazelle1984} are typical examples of PLCs that do not have a CDT. A sufficient condition for a PLC to admit a CDT is that all its \rev{segments} are \emph{strongly} Delaunay (cf. CDT theorem \cite{Shewchuk1998EThm}): a 
\rev{segment} is strongly Delaunay if it admits a circumsphere that neither contains nor touches other PLC vertices.
Interestingly enough, this characteristic deals with \rev{segments} only; therefore, splitting PLC \rev{segments} at appropriate points is sufficient to turn a generic PLC into one that admits a CDT. The split points are denoted as Steiner points. We thus allow such splitting for PLCs that do not admit CDTs. Note that splitting is a purely topological operation that does not modify the geometric realization of the input. Among the possible approaches to compute Steiner points, we chose the method used by Si \cite{HSi2005CDT} in his \texttt{tetgen} software, as it produces the smallest number of Steiner points in practice.

\subsection{Segment recovery} \label{sec:segmentrecovery}
Si observes that if all the \rev{segments} of a PLC belong to the Delaunay tetrahedrization of its vertices, then a CDT exists and can be computed using local operations (see \cite{HSi2005CDT}, thm. 2).
When a PLC \rev{segment} is not in the Delaunay tetrahedrization, it is called a \emph{missing} \rev{segment} and cannot be strongly Delaunay.
In this case, the idea is to split it so that the resulting (shorter) \rev{sub-segments} have more chances to satisfy the condition.
The exact location of the split is crucial to guarantee termination.

To determine split points, Si uses the concept of \emph{encroaching point}.
Let $e = \langle v_1, v_2\rangle $ be a missing \rev{segment}, $D$ be the smallest (diametral) sphere by $v_1$ and $v_2$, and $V_D$ be the set of vertices enclosed in $D$, excluding $v_1$ and $v_2$. Note that $V_D$ cannot be empty because $e$ would be strongly Delaunay and not missing. The vertices in $V_D$ are called \emph{encroaching points} for $e$. The algorithm picks the encroaching point $r$ for which the circle defined by $v_1$, $v_2$ and $r$ has the maximum radius, and declares it to be the \emph{reference} point for the \rev{segment}.

Another important concept in Si's approach is the classification of an \emph{acute} vertex: a vertex $v$ is acute if two PLC \rev{segments} incident at $v$ form an acute angle. Based on this, PLC \rev{segments} are classified into three categories:
\begin{enumerate}
\item \rev{Segments} having no acute vertices;
\item \rev{Segments} having only one acute vertex;
\item \rev{Segments} having two acute vertices.
\end{enumerate}


Missing \rev{segments} of the $3^{rd}$ category are split at their midpoint, so that the resulting sub-\rev{segments} belong to the $2^{nd}$ category. Any other missing (sub)\rev{segment} $\langle v_1, v_2\rangle$ with reference point $r$ is further subdivided by inserting a point $v_m$, until no \rev{segment} is missing. 




If $e$ is of the $1^{st}$ category, Si's method considers the two spheres $S_1$ and $S_2$ centered at $v_1$ and $v_2$ respectively, and both touching $r$.  If the radii of $S_1$ and $S_2$ are both bigger than half of the length of $e$, $v_m$ is set to the midpoint of $e$; otherwise, $v_m$ is the intersection of $e$ and the sphere that has the smallest radius (see inset).
\setlength{\intextsep}{1pt} 
\setlength{\columnsep}{4pt} 
\begin{wrapfigure}{r}{0.4\linewidth}
  \begin{center}
   \includegraphics[width=\linewidth]{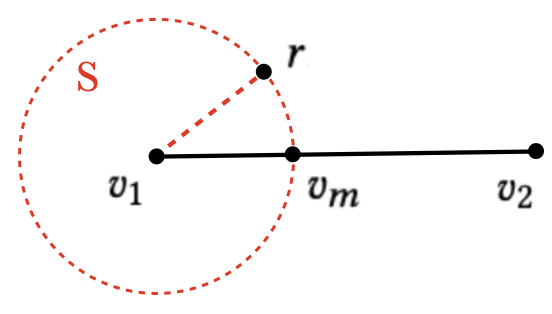}
  \end{center}
\end{wrapfigure}

If $e$ is in the $2^{nd}$ category and $w$ is its acute vertex, all its sub-\rev{segments} remain in the same $2^{nd}$ category and \emph{remember} the original acute vertex $w$.

$w$ is used as center for a sphere $S$ which touches the reference point $r$. Let $v_w$ be the endpoint closest to $w$, $v_o$ be the other endpoint, and $p$ be the intersection point of $S$ and $e$. If $p$ is closer to $r$ than to $v_o$, $v_m$ is set to $p$. 
Otherwise, if the distance $d$ between $r$ and $p$ is less than half the distance between $v_w$ and $p$, shift $p$ towards $v_w$ by a distance $d$. Otherwise, move $p$ to the midpoint of the segment $<v_w, p>$. $v_m$ is set to $p$.

Missing \rev{segments} are split one after the other in an iterative process which is guaranteed to converge to a PLC that admits a CDT.

Note that $v_m$ may be determined by the intersection of \rev{a segment} with a sphere, and hence its coordinates can be irrational numbers even if all the input vertices are in floating point precision.

\subsection{Face recovery}
\label{sec:facerecovery}
\begin{figure}
    \centering
    \includegraphics[width=\columnwidth]{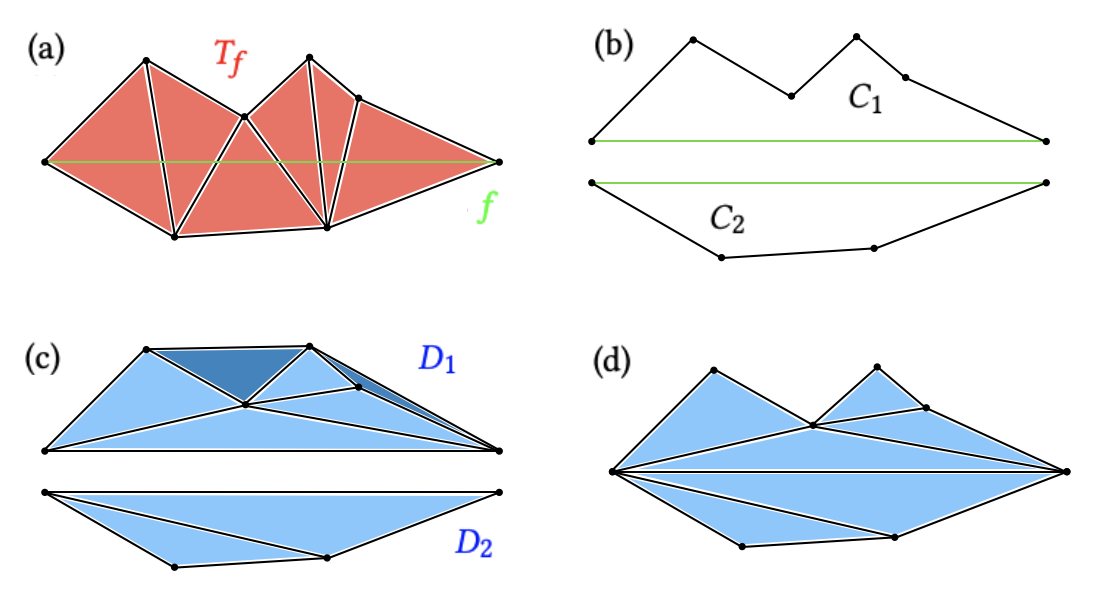}
    \caption{Face recovery in 2D. The missing PLC 2D-face $f$ (green) is intersected by mesh triangles $T_f$ (red)(a), which are removed to create two half-cavities (b). The convex hull of each half-cavity is then Delaunay-triangulated (blue) (c). Triangles outside the cavity (dark blue) are removed whereas the remaining triangles are used to fill the cavity while preserving $f$ (d).}
    \label{fig:facerec_flow_2D}
\end{figure}

Once all the PLC \rev{segments} are part of the Delaunay tetrahedrization, Si's method verifies whether all PLC faces are represented by the union of Delaunay triangles. A face that is not represented is denoted as a \emph{missing} face that must be \emph{recovered}. That means that the  region around the face must be retetrahedrized in order to enforce the requirement. A PLC-face $f$ is determined to be missing if it is pierced by at least one edge of the tetrahedrization. Let $T_f$ be the union of all the tetrahedra that are incident to these face-piercing edges. Each face-connected subset of $T_f$ forms a \emph{cavity} (see Fig. \ref{fig:facerec_flow_2D}) that requires retetrahedrization (two tetrahedra are face-connected if they share a triangle). 

The cavity is split along the plane of $f$ into two half-cavities, $C_1$ and $C_2$, where the vertices that belong to the plane of $f$ are assigned to both cavities. For each of the $C_i$, the vertices are used to compute a local Delaunay tetrahedrization $D_i$, which is used to fill the respective half cavity with new tetrahedra. Note that $D_i$ is convex but $C_i$ may be concave, meaning that not all the tets in $D_i$ are necessarily used. Then, the final retetrahedrization of the cavity is the union of the used tets from $D_1$ and $D_2$. All missing faces are recovered one after the other in an iterative process.

In order for this method to work, the triangles bounding $C_i$ should result in triangles in $D_i$. If a triangle $\tau \in C_i$ is not in $D_i$, the half-cavity is expanded by adding the opposite tetrahedron to $\tau$. In that case,  $C_i$ is updated and $D_i$ is recomputed. This process is iterated as long as the boundary of $C_i$ is not entirely represented by triangles in $D_i$. This might result in a failure of the algorithm, as we point out in the following.

\subsection{Possible failures in 
\texttt{tetgen}} \label{sec:issues}
While efficient, Si's method has two potential weaknessess. First, as stated in Sec. \ref{sec:segmentrecovery}, Steiner points may have irrational coordinates, meaning that a correct implementation requires predicates and data types that can robustly deal with irrational numbers (see Sec. \ref{sec:exact-cdt}). However, the only existing implementation (\texttt{tetgen}) uses floating-point arithmetics. Even if filtered exact predicates \cite{shew97_pred} are employed in \texttt{tetgen}, their guarantees are lost as soon as a Steiner point is rounded to its closest floating point representable position.

\begin{figure}
    \centering
    \includegraphics[width=\columnwidth]{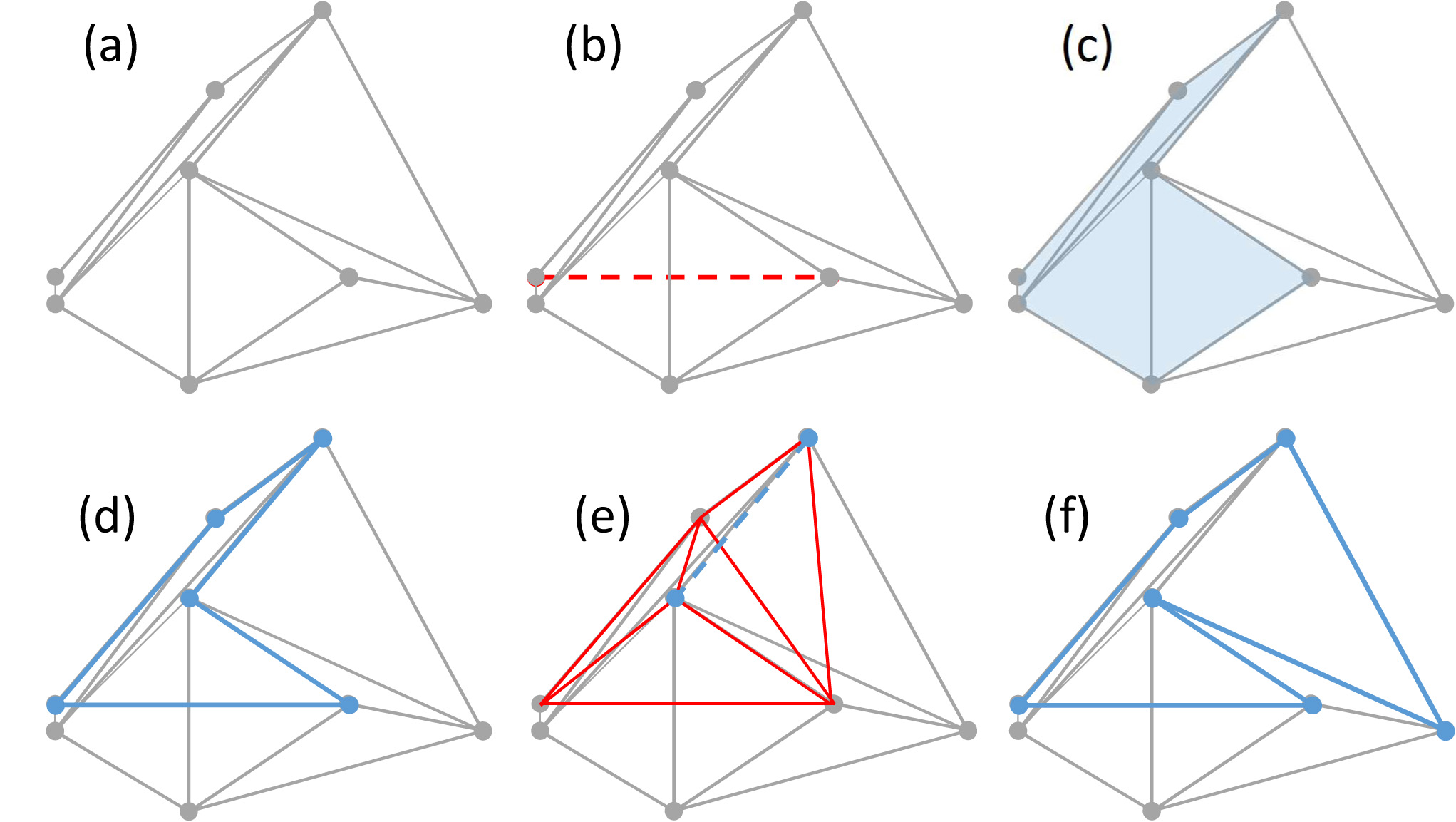}
    \caption{A 2D example showing a failure of the cavity expansion described in \cite{HSi2005CDT}, where 3D face recovery corresponds to 2D segment recovery. We start from the configuration in (a), which is plausible since having recovered any segment before, the Delaunay property of the vertices may not hold anymore. Consider the missing segment being recovered (b), its corresponding cavity (c), and the upper half cavity (d). The Delaunay triangulation of the vertices of the half-cavity  is in red (e), where the dashed blue \rev{segment} is missing, and thus the algorithm expands across it (f). However, this results in a half-cavity that intersects with the lower half-space.}
    \label{fig:expansion_fail}
\end{figure}

Second, the cavity expansion process described in Sec. \ref{sec:facerecovery} may fail regardless of numerics. The algorithm implicitly assumes that the local tetrahedrizations of the two half-cavities have disjoint interiors, and that their boundaries match at the face being recovered. However, there are cases where the expansion process adds a tetrahedron from across the plane of the recovered face (see Fig. \ref{fig:expansion_fail}), leading to an intersection of the two tetrahedrizations. Such cases are rare (this happens in just 2 of the 4408 models we tested). However, this is an algorithmic failure that cannot be overlooked.

\section{Robust CDT}
\label{sec:exact-cdt}


As mentioned in the introduction, representing Steiner points using floating point coordinates introduces a major weakness in any possible implementation. Employing thresholds does not solve the inherent problem, and might still lead to failure (Sec. \ref{sec:result_and_discussion}). 

In the following, we describe the individual components of our robust CDT algorithm. They comprise (1) a standard Delaunay tetrahedrization of the input vertices based on incremental insertion;
(2) a robust segment recovery procedure;
(3) a robust face recovery phase;
(4) an internal/external characterization of the tetrahedra.
While for (1) and (4) we adopt standard methods, our novel contributions reside in steps (2) and (3).
Our segment recovery proceeds as in \cite{HSi2005CDT}, but we resolve its numerical fragility thanks to our new implicit point type described in the remainder.
Similarly, our face recovery also proceeds as in \cite{HSi2005CDT} while employing our novel numerical kernel. Nonetheless, we also detect and resolve the cavity expansion theoretical issue and, in these cases, we proceed with an alternative algorithm (Sec. \ref{sec:giftwrap}), so as to guarantee a successful termination in all circumstances.


\subsection{\text{tetgen} with an exact number type}
\label{sec:exact-implementation}
The easiest solution to the fragility problem is to reimplement Si's approach with \emph{exact} arithmetic kernels. Nevertheless, one then needs the number type to be closed under the square-root operation. As an example, Steiner points may be at the intersection of a segment and a sphere (Sec. \ref{sec:segmentrecovery}), where then their coordinates could be irrational.

An appropriate library for that task is CORE~\cite{core1999}. CORE provides a number type that is closed under all the standard arithmetic operations, including the square root. CORE is based on the representation of intervals, and is able to evaluate the sign of expressions with guaranteed correctness. This however comes at a rather high computational cost. Even while exploiting the lazy evaluation paradigm provided by CGAL \cite{fabri2009cgal} to speed up calculations based on CORE, the resulting algorithm is far too slow for practical use (hours are necessary even for moderately small input files, see Sec. \ref{sec:result_and_discussion}). We note that this implementation is numerically exact, but still does not solve the theoretical flaw defined in Sec.~\ref{sec:issues}.

\subsection{Rational Steiner points}
\label{sec:nexact-implementation}
To mitigate the performance problem with CORE, we can employ a simple fix that allows us to work with (arbitrarily precise)  \emph{rational} numbers. We observe that Steiner points must be placed exactly on the \rev{segment} they split. If their exact position is \emph{snapped} to an approximated location (e.g., due to the need to represent coordinates using floating point numbers) we are actually deforming the input PLC. In turn, this renders all the theoretical proofs invalid and leads to non-robust implementations that can easily crash. Nevertheless, as long as a Steiner point remains exactly on its originating \rev{segment}, its position along the \rev{segment} can change up to a certain amount without causing a change in the mesh connectivity. We then opt to represent the approximate Steiner point with rational coordinates that are arbitrarily close to the exact location defined by Si's algorithm (Sec. \ref{sec:segmentrecovery}). Specifically, we parameterize this position by a linear combination of the \rev{segment} endpoints with a rational $t \in (0,1)$, so that the position $p$ is $t v_1 + (1-t) v_2$.
This allows reimplementing the algorithm using faster number types such as GNU GMP rationals \cite{gnugmp}. The value of $t$ can be approximated by rounding its exact (possibly irrational) value to its nearest floating point number using the standard \texttt{sqrt} library function. If this is not precise enough, the precision can be iteratively doubled as long as necessary.
However, our experiments reveal that in practice there is no actual need for such an increase in precision: using the initial approximation always leads to exactly the same results we could obtain using the exact implementation based on CORE. Even in this case, CGAL's lazy evaluation mechanism can be exploited to speed up the operations of a factor of $1.5\times$-$4\times$ (Sec. \ref{subsec:results}).

\subsection{Implicit Steiner points}
\label{sec:fast-implementation}
The rational version of our algorithm opens the door to a further, and major, optimization: we can replace GMP with custom indirect geometric predicates \cite{MAtteneIndPred}.

In a nutshell, indirect predicates use standard floating point calculations to derive \textbf{exact} information regarding the mutual position of so-called \emph{implicit} points. In turn, an implicit point is an unevaluated expression representing a position in space as a function of other known positions. In 3D, examples of implicit points are LPI (Line-Plane Intersection) and TPI (Three-Planes Intersection). When its floating point coordinates are known, a point may be called an \emph{explicit} point. With this terminology, an LPI is a function of five explicit points, two representing the line and three representing the plane being intersected.
LPIs and TPIs, along with standard \texttt{orient2D} and \texttt{orient3D} geometric predicates operating on them, were successfully used for diverse applications \cite{cherchi2020, DiazziAttene2021, du2022}.
Herewith, we introduce a new type of implicit point representing a linear combination of two known points, and show how to derive indirect versions of the \texttt{orient3d} and \texttt{inSphere} predicates necessary to construct a CDT.

Let $v_1$ and $v_2$ be two explicit points, and let $t$ be a floating point number in the range $(0, 1)$. The expression $t v_1 + (1-t) v_2$ is called an LNC (LiNear Combination) implicit point. LNCs can be effectively used to represent Steiner points during all the phases of our CDT algorithm. Indeed, any non-explicit point treated in our approach is a point that subdivides an input segment, meaning that it can be expressed as a linear combination of its two (explicit) endpoints.
Note that, thanks to our \emph{local-global} approach to exploit symbolic perturbation (Sec. \ref{sec:giftwrap}), we do not need any additional points to remove the so-called \emph{local degeneracies} discussed in Sec. 6 of \cite{HSi2005CDT}.

\subsubsection{Indirect \texttt{orient3d} and \texttt{inSphere} predicates}
\label{sec:indirect_preds}

The basic idea behind indirect predicates is to combine the expression of the predicate itself with the expression of the argument points. If the combined expression is either a polynomial or a ratio of polynomials, its sign can be determined without errors using arithmetic filtering and, if necessary, floating point expansions \cite{MAtteneIndPred}.

In our context, argument points can be: (1) explicit points whose coordinates come directly by reading an input file or (2) implicit LNCs representing Steiner points.
In either case, the point expression is a simple polynomial, and the result of \texttt{orient3d($p_1, p_2, p_3, p_4$)}, where all the four argument points are explicit, corresponds to the sign of the following determinant, which is another polynomial:
\begin{align}
\label{eq:orient3d}
\begin{vmatrix}
p_{1x} - p_{4x} & p_{1y} - p_{4y} & p_{1z} - p_{4z} \\
p_{2x} - p_{4x} & p_{2y} - p_{4y} & p_{2z} - p_{4z} \\
p_{3x} - p_{4x} & p_{3y} - p_{4y} & p_{3z} - p_{4z}
\end{vmatrix}
\end{align}
If the first argument point  is an LNC $p_1 = t v_1 + (1-t) v_2$, while the other three points are explicit, the combined expression can be obtained by simply replacing $p_{1x}$ with $t v_{1x} + (1-t) v_{2x}$ (and similarly for $p_{1y}$ and $p_{1z}$). Therefore, the corresponding indirect \texttt{orient3d} predicate evaluates as:
\begin{small}
\begin{align}
\label{eq:orient3dlnc}
\begin{vmatrix}
t v_{1x} + (1-t) v_{2x} - p_{4x} & t v_{1y} + (1-t) v_{2y} - p_{4y} & t v_{1z} + (1-t) v_{2z} - p_{4z} \\
p_{2x} - p_{4x} & p_{2y} - p_{4y} & p_{2z} - p_{4z} \\
p_{3x} - p_{4x} & p_{3y} - p_{4y} & p_{3z} - p_{4z}
\end{vmatrix}
\end{align}
\end{small}
Similarly, the result of \texttt{inSphere($p_1, p_2, p_3, p_4, p_5$)}, where all the five argument points are explicit, corresponds to the sign of the following determinant:
\begin{align}
\label{eq:insphere}
\begin{vmatrix}
p_{1x} - p_{5x} & p_{1y} - p_{5y} & p_{1z} - p_{5z} & \lVert p_1 - p_5 \rVert ^2 \\
p_{2x} - p_{5x} & p_{2y} - p_{5y} & p_{2z} - p_{5z} & \lVert p_2 - p_5 \rVert ^2 \\
p_{3x} - p_{5x} & p_{3y} - p_{5y} & p_{3z} - p_{5z} & \lVert p_3 - p_5 \rVert ^2 \\
p_{4x} - p_{5x} & p_{4y} - p_{5y} & p_{4z} - p_{5z} & \lVert p_4 - p_5 \rVert ^2
\end{vmatrix}
\end{align}
The matrix above can be modified by simple substitution as done for \texttt{orient3d} to derive indirect predicates. 

In principle, any combination of explicit and implicit points determines one different indirect predicate, meaning that we need to account for \rev{16} different versions of the \texttt{orient3d} predicate and \rev{32} versions of the \texttt{inSphere}. Fortunately, since swapping rows in the matrix affects the determinant sign in a predictable manner, one can limit the possible variety to 5 and 6 versions respectively. One version accounts for the total number of argument points that are in implicit form and assumes these points are the first in the argument list.
When calling the predicate, the code swaps contiguous rows to move all implicit points to the beginning of the list and keeps track of the parity of these swaps to determine the sign. 
For example, if the only implicit point is $p_2$, the result of \texttt{inSphere($p_1, p_2, p_3, p_4, p_5$)} is computed as \texttt{-inSphere($p_2, p_1, p_3, p_4, p_5$)}, and the minus sign is there because the number of necessary swaps is odd. If $p_2$ and $p_3$ are implicit, \texttt{inSphere($p_1, p_2, p_3, p_4, p_5$)} is equal to \texttt{inSphere($p_2, p_3, p_1, p_4, p_5$)}, because the number of swaps is even.
Filter values for fast calculation using floating point arithmetic are given in Appendix \ref{ap:filters}.

\subsection{Modified gift-wrapping algorithm}
\label{sec:giftwrap}
To cope with the cavity expansion failures, we describe an alternative algorithm based on \cite{Shew00GW}. 
Shewchuk's method is guaranteed to produce the CDT out of a PLC that admits one. It is essentially a modification of a na\"ive gift-wrapping approach: first, each face in the input PLC is triangulated using a local 2D CDT. Then, each resulting triangle is connected to one \emph{apex} vertex to form a tetrahedron. If the input vertices are in general position, the CDT is unique, meaning that only one vertex in the set is a valid apex for any given triangle. Hence, when such a valid apex is selected among the input vertices, the tetrahedron is guaranteed to be part of the eventual CDT.

In our method, we combine a local version of this gift-wrapping algorithm with a coherent symbolic perturbation technique \cite{symbolicperturbation} that guarantees conformity everywhere, even if the points are not in general position. Specifically, we construct one cavity at a time as described in Sec. \ref{sec:facerecovery} and split it into two half-cavities. For each half-cavity, we build one tetrahedron at a time as in \cite{Shew00GW} while considering three key aspects:
\begin{itemize}
\item The triangles common to the opposite half-cavity are unknown when tetrahedrizing the first of the two halves;
\item The triangulation induced by the tetrahedrization of the two half-cavities must match on the common face, even if the vertices are not in general position;
\item The implementation may not tolerate numerical errors, and therefore all predicates and checks must be exact.
\end{itemize}

For any missing PLC-face $f$ we first delete all tetrahedra $T_f$ whose interior intersects $f$ and keep track of all their vertices $V_f$. When done, we keep track of all the triangles $\partial C$ that bound the resulting cavity $C$, each oriented so that the normal points toward the exterior.
Then, we create the half-cavities $C_1$ and $C_2$ by splitting $C$ through $f$ and, at the same time, we split $\partial C$ in two subsets $\partial C_1$ and $\partial C_2$. Furthermore, we split the set $V_f$ in two subsets $V_1$ and $V_2$ whose vertices are over (or on) and below (or on) $f$ respectively.

We then fill one half-cavity at a time.
Let $C_1$ be the first. The set $\partial C_1$ contains \emph{known} triangles that bound $C_1$, but does not completely enclose the half-cavity. Nonetheless, we have sufficient information to proceed with the creation of tetrahedra.


During our iterative process, a set of triangles $\partial C_{cur}$ defines the \emph{current} boundary of the half-cavity being filled. At the beginning, $\partial C_{cur} = \partial C_1$.

\begin{figure}
    \centering
    \includegraphics[width=\columnwidth]{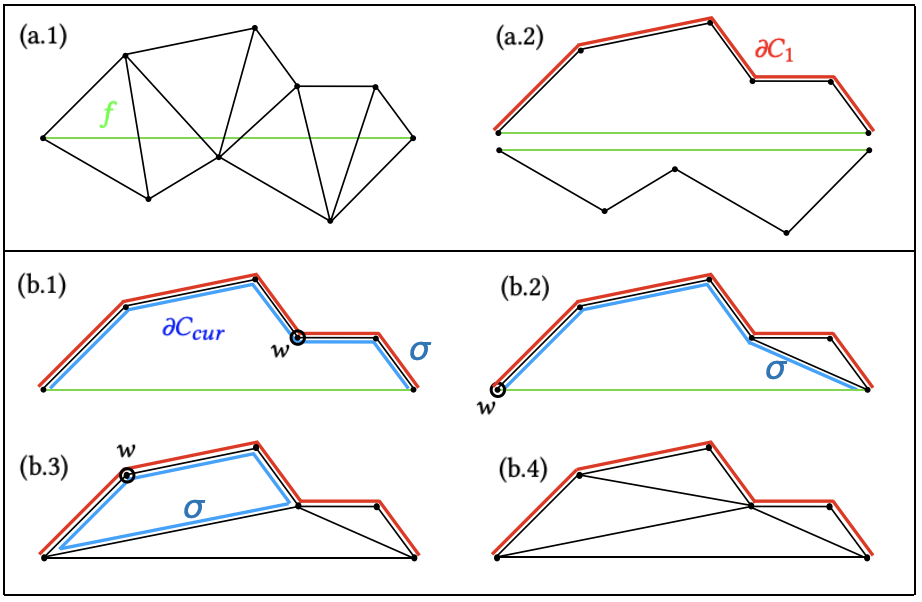}
    \caption{A 2D example of gift wrapping. The missing PLC \rev{segment} $f$ (green) is intersected by mesh triangles (a.1), which are removed to create two half-cavities (a.2). The upper cavity is then triangulated (b.1-b.4), by exploiting gift-wrapping algorithm discussed in Sec. \ref{sec:giftwrap}. The cavity boundary $\partial C_1$ (red) does not change during triangulation, while the current cavity boundary $\partial C_{curr}$ (blue) does. At each step an \rev{edge} $\sigma$ of $\partial C_{curr}$ is connected to an apex $w$ (circled) to create a \emph{valid} triangle. }
    \label{fig:gw_2Dflow}
\end{figure}

We iteratively pick a triangle $\sigma$ from $\partial C_{cur}$ and search for a suitable apex vertex $w$ in $V_1$ such that the resulting tetrahedron $t$ is \emph{valid}. When we find it, we create the tetrahedron, update $\partial C_{cur}$ and process the next $\sigma$. After the first iteration, $\sigma$ might be on the plane of $f$: in this case we simply skip it and move to the next one. The process terminates when all the triangles in $\partial C_{cur}$ are processed, \rev{see Fig. \ref{fig:gw_2Dflow}}.

During this process the tetrahedron $t$, made by joining the triangle $\sigma$ and the apex $w$, is valid if it satisfies all the following conditions: 
\begin{itemize}
    \item[i)] $w$ is in the opposite half-space with respect to $\sigma$ outgoing normal (i.e. $t$ has positive volume);
    \item[ii)] if $t$ intersects a triangle of $\partial C_{cur}$ then the intersection is a common subsimplex;
    \item[iii)] no vertex in $V_1$ is in the circumsphere of $t$, except those whose visibility is occluded by $\partial C_1$.
\end{itemize}

We observe that, as soon as a vertex in $V_1$ is no longer \emph{usable} as an apex (e.g., because, as the wrapping proceeds, it becomes completely surrounded by tetrahedra) it might be removed from the search list. However, condition $iii)$ must still check all the vertices in the original half-cavity.

When $C_1$ is completed, we repeat the same process on $C_2$, though in this case we reuse the triangles produced on $f$ while filling $C_1$.

Note that, in the case of cospherical points in the cavity, the CDT may be not  unique. This means that, when creating a tetrahedron for $C_2$, we are no longer sure that it will eventually induce a mesh that conforms with the triangles on $f$ inherited from $C_1$. Avoiding the inheritance (and hence proceeding for $C_2$ with an open boundary as we do for $C_1$) would not solve the problem, because the common triangulation induced on $f$ might not match.

We solve this problem by exploiting symbolic perturbation \cite{symbolicperturbation} as follows: let $v_1, ..., v_4$ be the four vertices of a tetrahedron, and let $q$ be a query vertex. Our exact \texttt{inSphere} predicate (see Sect. \ref{sec:indirect_preds}) states whether $q$ is \emph{inside}, \emph{outside}, or \emph{exactly on} the sphere defined by $v_1, ..., v_4$. When the result is \emph{exactly on}, we take a decision between \emph{inside} and \emph{outside} based on the order in which the five vertices are stored in memory (Alg. \ref{alg:pinsphere}).
To guarantee that all the cavities are tetrahedrized conformally with the other parts of the mesh, we create the subvectors representing $V_1$ and $V_2$ so that any two vertices in $V_1$ (resp. $V_2$) are stored in the same order as they are stored in the \emph{global} mesh vector.

\begin{algorithm}

\caption{\rev{perturbedInCircumphere}($i_1, i_2, i_3, i_4, i_5$)}
\label{alg:pinsphere}

\textbf{Input:}
$i_1, ..., i_5$: indexes of the five points in a global vector $V$, \rev{where $i_1, ..., i_4$ are the four vertices of a valid tetrahedron, whereas $i_5$ is a query point} \\
\textbf{Output:} \rev{-1 if $i_5$ is inside or on the circumsphere of $i_1, ..., i_4$, 1 if it is outside or on the circumsphere.}\\

\hrulefill \\
\SetAlgoLined
\vspace{0.1em}

r := inSphere($V[i_1], V[i_2], V[i_3], V[i_4], V[i_5]$)

\lIf{ $r \neq 0$ }{ \textbf{return r}  } 

sort $i_1, ..., i_5$ in ascending order by $n$ iterative swaps\\

r := orient3D($V[i_2], V[i_3], V[i_4], V[i_5]$)

\lIf{ $n$ is odd }{ r := -r  }

\lIf{ $r \neq 0$ }{ \textbf{return r}  }

r := orient3D($V[i_1], V[i_3], V[i_4], V[i_5]$)

\lIf{ $n$ is even }{ r := -r  }

\textbf{return r}

\end{algorithm}

We use indirect predicates for all the intersection and visibility checks required by conditions $(i)-(iii)$, so as to guarantee exactness without the need for slow exact arithmetic. Checking conditions $(ii)$ and $(iii)$ is particularly complex, as one must take into account all the possible configurations in which a triangle and a tetrahedron may be arranged in three-dimensional space. We describe these conditions in terms of geometric predicates in App. \ref{app:GWpred}.

\rev{The fact that this process leads to a CDT is given for granted in \cite{Shewchuk2002CDT} but may be not obvious at a first sight. We observe that a tetrahedrization is constrained Delaunay if and only if all the internal triangles are \emph{locally} constrained Delaunay. In turn, an internal triangle is locally constrained Delaunay if (1) it is a constraint (i.e. part of the input PLC) or (2) each of its two incident tetrahedra has a circumsphere that does not contain the apex vertex of the opposite incident tetrahedron. This precondition holds before digging the cavity, and still holds for all the triangles created within the cavity because the new tetrahedra are \emph{valid}. Each triangle on $\partial{C}$ is also locally constrained Delaunay because it must satisfy one of the following two conditions: (1) it is a constraint or (2) its incident tetrahedron outside the cavity does not contain any visible vertex (because the mesh was a CDT before digging the cavity) thus including the apex of its (newly created) opposite tetrahedron. Hence, the resulting mesh after gift-wrapping is a CDT wrt all the facets recovered so far.}

\begin{figure}
    \centering
    \includegraphics[width=\columnwidth]{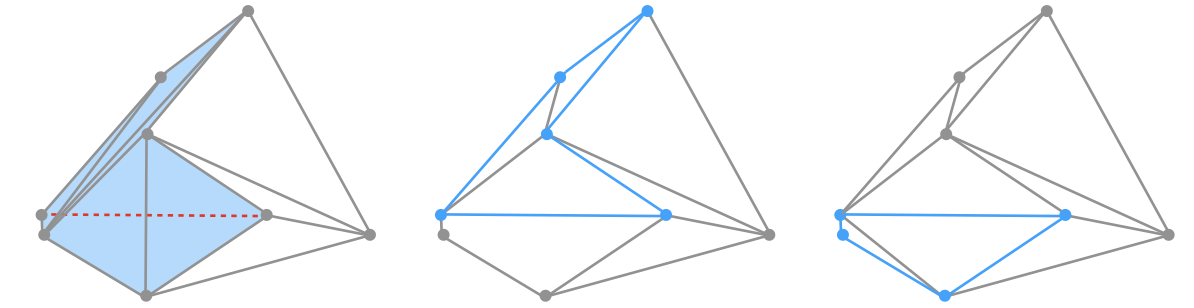}
    \caption{\rev{If the half-cavity expansion fails as for the 2D example on the left (see Fig. \ref{fig:expansion_fail}), we switch and fill the original (unexpanded) half-cavity with valid \emph{triangles} using gift-wrapping (middle), we then use it again for the other half-cavity (right).}}
    \label{fig:gw_resolved}
\end{figure}

On average, our modified gift-wrapping algorithm is fast enough and allows us to successfully tetrahedrize all the models in our reference dataset (Sec. \ref{sec:result_and_discussion}). Nevertheless, since the approach in \cite{HSi2005CDT} is faster in practice, our implementation uses the latter unless the cavity expansion fails (Sec. \ref{sec:issues}), where then it switches to the modified gift-wrapping approach\rev{, see Fig. \ref{fig:gw_resolved}}. Note that cavity expansion fails in 2 out of 4408 models in our dataset.

We observe that, as an alternative to our modified gift-wrapping, cavities may be retetrahedrized  using iterative flips as in \cite{shewchuk2003updating}. However, this would require implementing further indirect predicates operating with four-dimensional points.

\subsection{Interior/exterior characterization}
If the input unambiguously separates the space into internal and external parts, we can similarly characterize each tet in our final tetrahedrization.
\begin{figure}
    \centering
    \includegraphics[width=\linewidth]{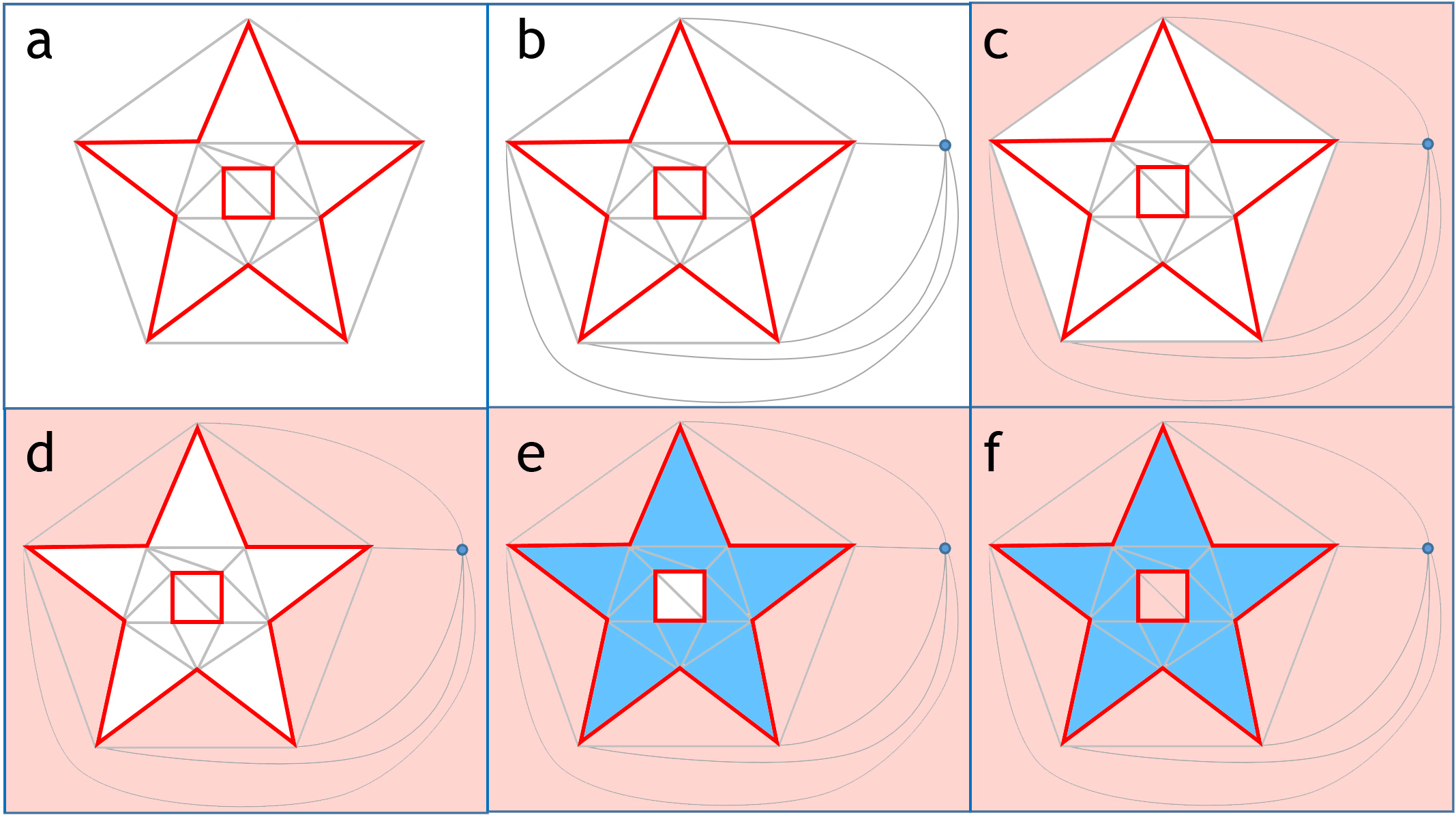}
    \caption{A 2D example of propagating internal (blue) and external (light red) labels. A PLC (red segments) is depicted along with its CDT (gray) (a). We add a ghost vertex (blue) (b), and connect it to all the boundary edges to form ghost \emph{triangles}. Ghost triangles are tagged as \emph{external} (c). The \emph{external} label propagates without crossing PLC edges (d), switches to \emph{internal} and propagates again (e), and finally switches back to \emph{external} and propagates on all the remaining triangles (f).}
    \label{fig:in_out}
\end{figure}
For this step, which is optional in our pipeline, we rely on a flood-filling approach previously adopted in other meshing algorithms \rev{\cite{Hu:2020, tetbool}}.
We use the common concept of \emph{ghost} vertex to ensure that each tetrahedron always has four neighbors. The ghost is essentially a virtual vertex ``at infinity'' that is connected to all the triangles of the convex hull, so that each triangle produces a \emph{ghost tetrahedron}. Adoption of ghosts significantly simplifies data management \cite{marot2019one}.

Our interior/exterior characterization proceeds by first assigning an \emph{external} label to all the ghost tetrahedra (Fig. \ref{fig:in_out}(c)), and then by propagating this label across \emph{unconstrained} triangular faces (i.e., faces which are not part of the input PLC, Fig. \ref{fig:in_out}(d)). When propagation stops (i.e., because all the faces reached are constrained), we switch the label from \emph{external} to \emph{internal} across PLC faces, and keep propagating across unconstrained faces (Fig. \ref{fig:in_out}(e)). If propagation stops on other constrained faces we switch the tag back to \emph{external} and so on, until all the tetrahedra are reached (Fig. \ref{fig:in_out}(f)).

This simple process assumes that the input defines a well-defined volume or, equivalently, that each \rev{segment} in the PLC has an even number of incident faces. If this is not the case one may decide to find a plausible characterization using winding numbers \cite{jacobson2013robust} or graph labeling \cite{DiazziAttene2021}, but this is out of the scope of this work.

\subsection{Implicit Steiner CDT algorithm}
\label{sec:steinercdt}
Having defined the individual steps, we next describe the entire algorithm which includes the possible refinement of the input PLC to make it admit an output CDT.
With reference to Alg. \ref{alg:steinercdt}, we first compute the Delaunay tetrahedrization $D$ of the input vertices using a classical Bowyer-Watson incremental insertion \rev{ \cite{bowyer81, watson81} }.
Then, we proceed with the segment recovery (Sec. \ref{sec:segmentrecovery}). This step is enclosed in a \textbf{while} loop because a non-missing \rev{segment} might become missing when $D$ is modified due to a Steiner point insertion. A proof of convergence is given in \cite{HSi2005CDT}.

The subsequent face recovery works similarly. Here the outer loop is necessary because a non-missing face might become missing due to the cavity expansion. Even in this case, proof of convergence is given in \cite{HSi2005CDT}. Note that gift-wrapping is used only if the expansion fails, which means it does not have an impact on guaranteed convergence.

\begin{algorithm}

\caption{SteinerCDT($P$)}
\label{alg:steinercdt}

\textbf{Input:}\\
$P = <V, E, F>$: a valid PLC with vertices $V$, edges $E$ and faces $F$.\\

\textbf{Output:}\\
A Steiner CDT of $P$

\hrulefill \\
\SetAlgoLined
\vspace{0.1em}

$D$ := Delaunay($V$) \tcp{\cite{bowyer81}}

\While{at least a \rev{segment} is missing in $D$}{
\ForEach{$e \in E$}{
 \If{$e$ is missing in $D$}{
  calculate a Steiner point $s$
  \tcp{Sec. \ref{sec:segmentrecovery} and \ref{sec:nexact-implementation}}
  \
  split $e$ at $s$\\
  insert $s$ in $D$  \tcp{\cite{bowyer81}}\
 }
}
}
\While{at least a face is missing in $D$}{
\ForEach{$f \in F$}{
create half-cavities $C_1$ and $C_2$
\tcp{Sec. \ref{sec:facerecovery} and \ref{sec:giftwrap}}\
\eIf{$C_1$ and $C_2$ can both be expanded}{
recover $f$ in $D$ by local Delaunay
 \tcp{Sec. \ref{sec:facerecovery}}
}
{
recover $f$ in $D$ by gift-wrapping
 \tcp{Sec. \ref{sec:giftwrap}}
}
}
}
\textbf{return} $D$
\end{algorithm}
\section{Post-processing and Applications}
\label{sec:applications}

\subsection{Floating point representation}
\label{sec:fp-representation}

\begin{figure*}
    \centering
    \includegraphics[width=\linewidth]{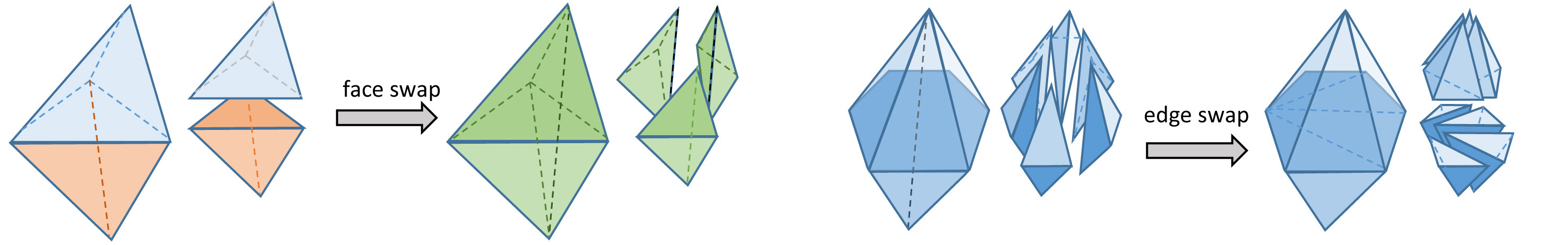}
    \caption{The two basic topological operators used to remove bad-shaped elements.}
    \label{fig:swaps}
\end{figure*}
Because the coordinates of LNC points can be losslessly converted to rational numbers, our implementation can save the output CDT to a file with no approximations.
However, a typical requirement in downstream applications is that the vertex coordinates are in floating-point precision, regardless of the number type used by the algorithm. Unfortunately, rounding our implicit points to their closest floating-point position may make a nearly degenerate (though valid) element into a flat or inverted tetrahedron.
In contrast to 2D Delaunay triangulations, the empty-sphere property does not always result in optimal element quality in 3D  \cite{alexa_harmonic}. In particular, 3D Delaunay meshes often exhibit bad-shaped elements called \emph{slivers}. A sliver is a valid tetrahedron with no short edges, but rather four nearly-coplanar vertices. Some methods remove these elements from a fully-Delaunay tetrahedrization (e.g.,\cite{sliver_exudation}); we however must constrain the output to exactly preserve the input PLC.  Though PLC-preserving sliver-removal algorithms have been designed (e.g., \cite{cheng_siam}), they are extremely complicated and refine the mesh everywhere, adding extra unnecessary elements.

Our approach is to remove bad-shaped tetrahedra by iterating local connectivity modifications so as to monotonically increase the element quality. While this is not a full mesh optimization such as in \rev{\cite{Hu:2018}}, we found it is sufficient to prevent the introduction of degenerate elements when rounding.

We modify the connectivity by iterative face and edge swaps. A face swap (also called a 2--3 swap) replaces two tetrahedra sharing a face $f$ with three tetrahedra sharing an edge that connects the two vertices opposite to $f$ in the initial configuration (see Fig.~\ref{fig:swaps}). An (3-2) edge swap is essentially the inverse operation to face swap, when applicable. Typical implementations use the said 3-2 swaps, 4-4 swaps, or even 5-6 swaps \rev{\cite{Hu:2018}}.  We use a single, but general, edge-swap operation that works as follows: we first split an edge $e = (v_1, v_2)$ by inserting a virtual point, then immediately collapse the new point to one of its neighbors different from $v_1$ and $v_2$. The temporary point is not assigned any position as it is immediately destroyed, hence the term \emph{virtual}.
An edge shared by $n$ tetrahedra can be swapped in $n-2$ different ways, depending on the neighboring vertex used to collapse the virtual point. When $n=3, 4 , 5$, our generic edge-swap corresponds to the standard 3-2, 4-4, and 5-6 swap respectively.

In our mesh improvement algorithm, each swap is operated only if both these conditions hold: (1) no tetrahedron is inverted or flattened due to the change; (2) the maximal AMIPS energy \rev{\cite{Hu:2018}} of tetrahedra strictly decreases due to the change.
When swapping an edge we randomly select a neighboring vertex to which we collapse the virtual point, where the conditions would hold; if no such vertex is found the swap is rejected. To ensure that the resulting mesh is still conformal with the input PLC, no swap is operated if the interior of the affected region contains constrained facets. After this process, the rounding may still introduce invalid tetrahedra, but in practice this possibility is dramatically reduced (Sec. \ref{sec:result_and_discussion}).
\rev{Clearly, after this process the mesh is no longer guaranteed to be constrained Delaunay.}
\rev{Note that, although an FP-rounded Steiner point is no longer \emph{exactly} on its originating input segment, that segment is unique and known, therefore the Steiner point can inherit boundary conditions from the input with no ambiguity.}

\subsection{Delaunay refinement}
\label{sec:delaunay-refinement}
\begin{figure}
    \centering
    \includegraphics[width=\linewidth]{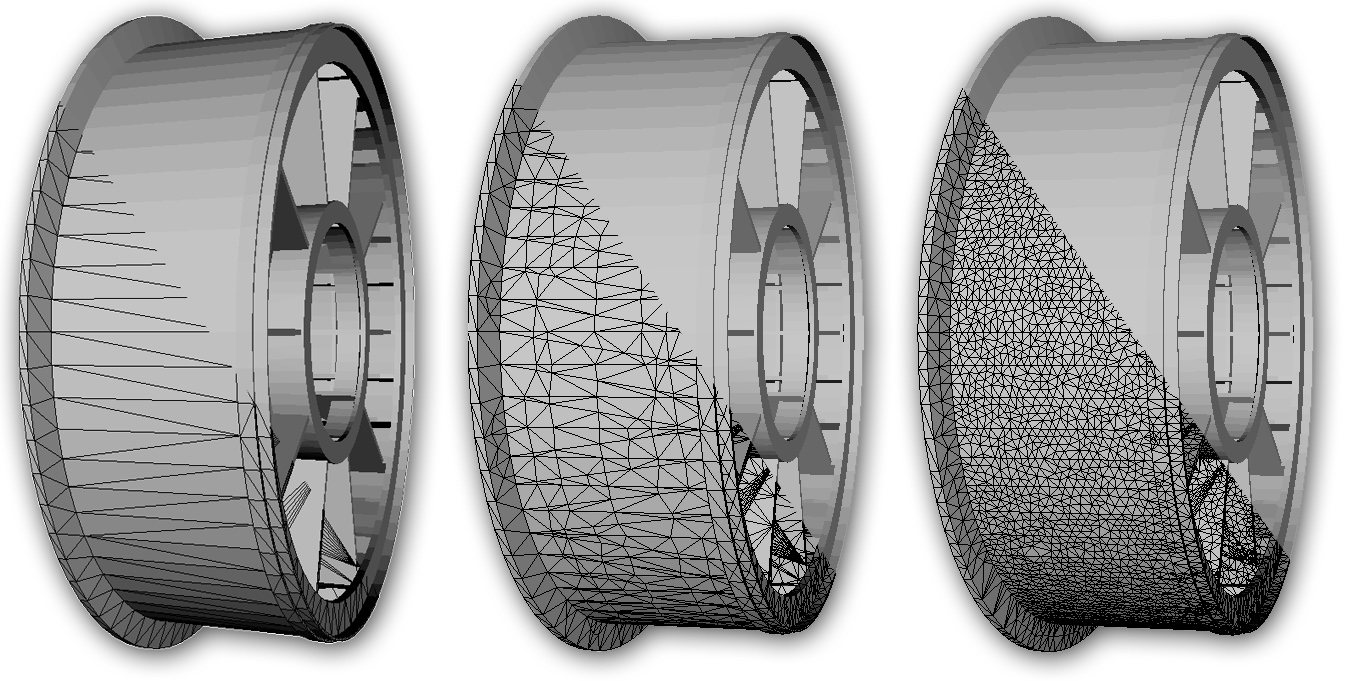}
    \caption{An input PLC that makes tetgen crash (left), admits a valid CDT by our method (middle), which is optimized by running tetgen in \emph{refine-only} mode on our CDT (right).}
    \label{fig:optim_tetgen}
\end{figure}

The CDTs produced by our method can be effectively used within a plethora of mesh refinement algorithms, each striving to maximize/minimize some particular metric depending on the target application. Most importantly, the fact that our meshes are \emph{constrained Delaunay} guarantees that Delaunay refinement
algorithms converge to reliably good meshes  \cite{shewchuk2000mesh, Shewchuk2002CDT}.
To demonstrate this property, we use the Delaunay refinement algorithm implemented in \texttt{tetgen} \cite{shewchuk2014higher}, where we set \texttt{tetgen} to bypass its initial CDT generation and instead provide the CDT from our algorithm as input. Fig. \ref{fig:optim_tetgen} shows an example.

If a certain amount of displacement is tolerable, other refinement algorithms may provide better results. We further tested our meshes as input to the algorithm in \cite{mmg_paper}, for which an open-source implementation is available \cite{mmglib}. We demonstrate this in Fig. \ref{fig:teaser}-(d, e).


\section{Results and discussion}
\label{sec:result_and_discussion}

We implemented our algorithm as a standalone tool in C++. For this, we modified the Indirect Predicates library  \cite{ipredslib} to support LNC points. We can plug in several number types to our code to represent coordinates (Secs. \ref{sec:exact-implementation}---\ref{sec:fast-implementation}). This enables a fair comparison of the various versions of the algorithm, in terms of performance and robustness, while being sure that exactly the same algorithm is run. We use the exact number types provided in CGAL \cite{fabri2009cgal} to implement the exact and rational versions. Our source code is freely available at \emph{https://github.com/MarcoAttene/CDT}. Our algorithm was run on a Linux-based machine equipped with an AMD EPYC 7452 CPU and 1Tb RAM. All the experiments were run on a single core.

\subsection{Results}
\label{subsec:results}

We tested our implementation on the meshes of the Thingi10k dataset \cite{Thingi10K} that satisfy the PLC conditions (i.e. that do not self-intersect), for a total of 4408 models. Our experiments demonstrate that our method is fast enough for most practical applications: 76\% of the models are processed in less than 1 second, and the average execution time is 4.3 seconds. Furthermore, only 1.2\% of the models require more than 60 seconds, in which the worst case takes 33 minutes. If one considers this small tail of models to represent outliers, the expected (average) time for the remaining 98.8\% of the models is just 1.8 seconds. We do not observe any clear relation between the number of input triangles and elapsed time. We argue that geometric characteristics of the input (i.e. the local feature size) are an overriding factor in this sense.

\begin{figure}
    \centering
    \includegraphics[width=0.8\columnwidth]{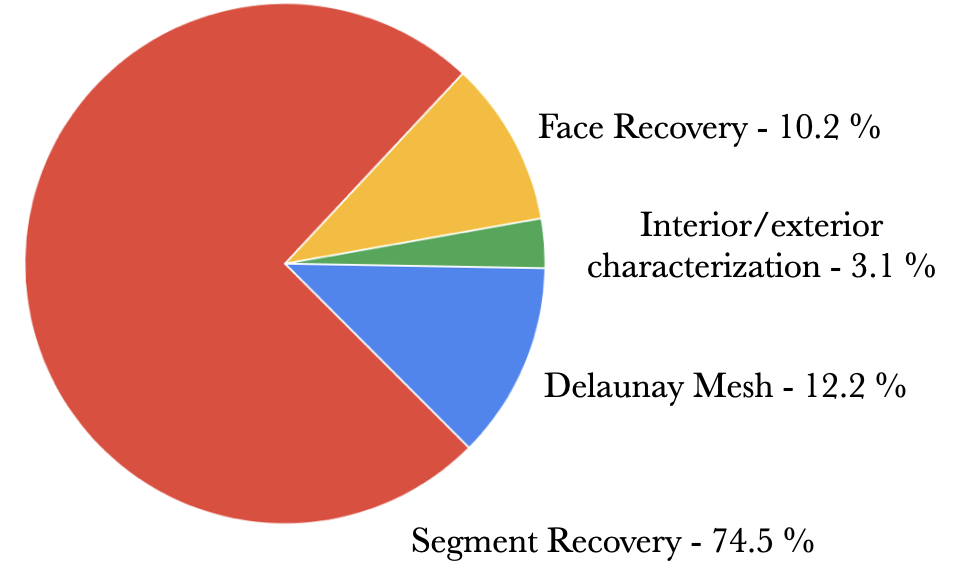}
    \caption{A division of our running time to the various steps of the pipeline. These data refer to the time spent processing the 4408 Thingi10k models defining a PLC.}
    \label{fig:pie-chart}
\end{figure}
\begin{figure}
    \centering
    \includegraphics[width=\columnwidth]{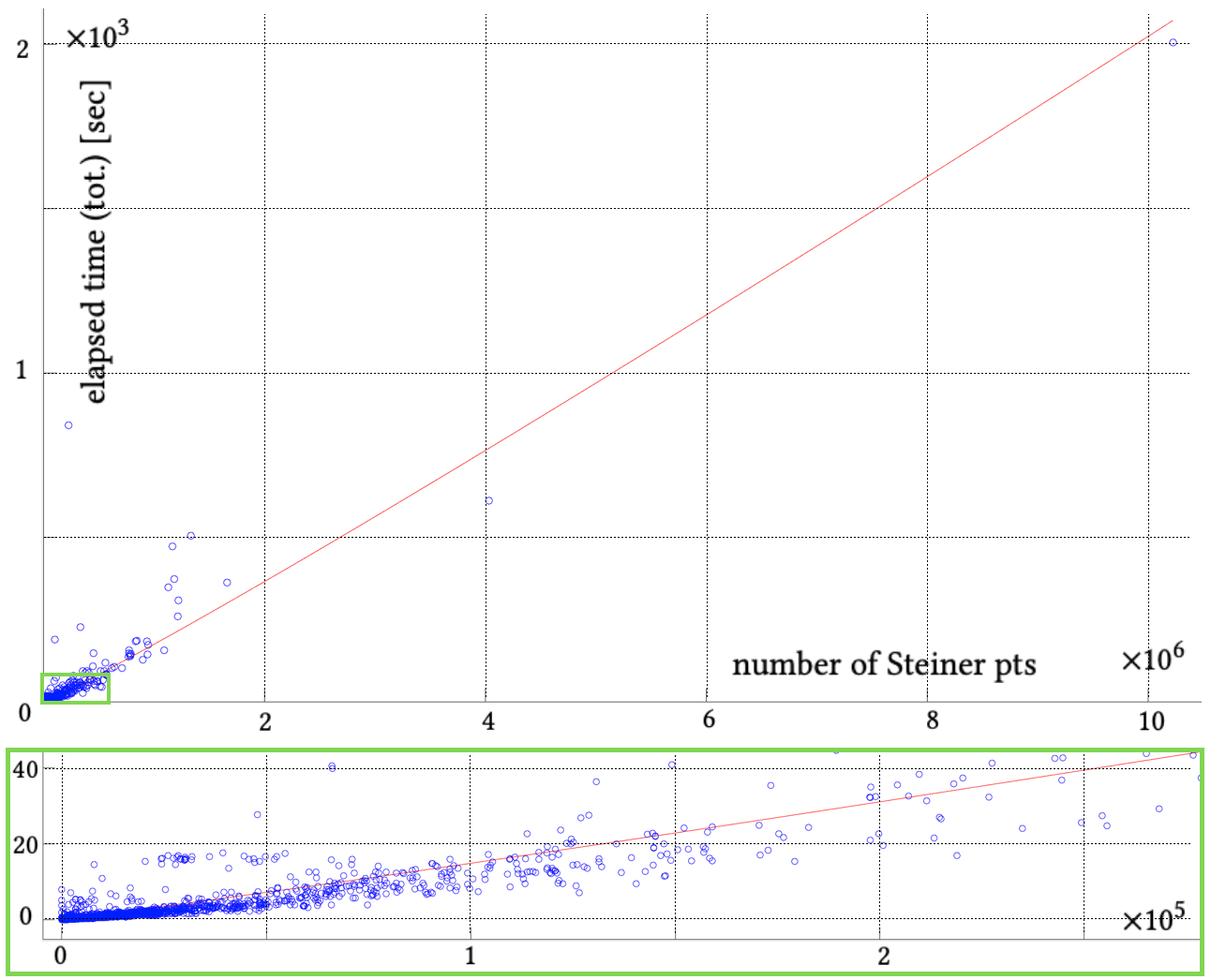}
    \caption{Elapsed time vs. number of Steiner points inserted. Each blue circle represents one of the 4408 models in our dataset. The red line is a least-square fit of a function of the form $t = A + B n \log(n)$, with $A=0.452$ and $B=0.0000125$. A zoom-in near the origin is shown in the bottom part.}
    \label{fig:stpVStime}
\end{figure}

Our algorithm may be divided into the following main phases:  Delaunay tetrahedrization of the input vertices, segment recovery, face recovery, and interior-exterior characterization of the tets. Fig. \ref{fig:pie-chart} gives an idea of how each of these phases impacts the overall execution time, and reveals that segment recovery accounts for 74.5\% of the total. This is an average value on the entire dataset, and clearly depends on the number of Steiner points required by each specific model (see Fig. \ref{fig:stpVStime}). For the model requiring the longest execution time, segment recovery takes 86\% of the total and adds 10216352 Steiner points. This shows where future research should focus to target further performance improvement (Sec. \ref{subsec:limitations}).

Memory usage is also kept within reasonable bounds and allows processing all the selected 4408 Thingi10k models even on an average home desktop PC.
On average, the peak memory allocated while processing a model is 46.5 Mb (as measured by \emph{getrusage()} \footnote{https://www.gnu.org/software/libc/manual/html\_node/Resource-Usage.html}), while worst case requires 8.88 Gb.

\rev{Our algorithm inserts less than 50k Steiner points in 91\% of our test dataset. About 25k Steiner points are used on average, with a worst case requiring 10 million points. The average decreases to 4.9k if the 9\% of models requiring more than 50k Steiner points are not considered.}

\begin{figure}
    \centering
    \includegraphics[width=\columnwidth]{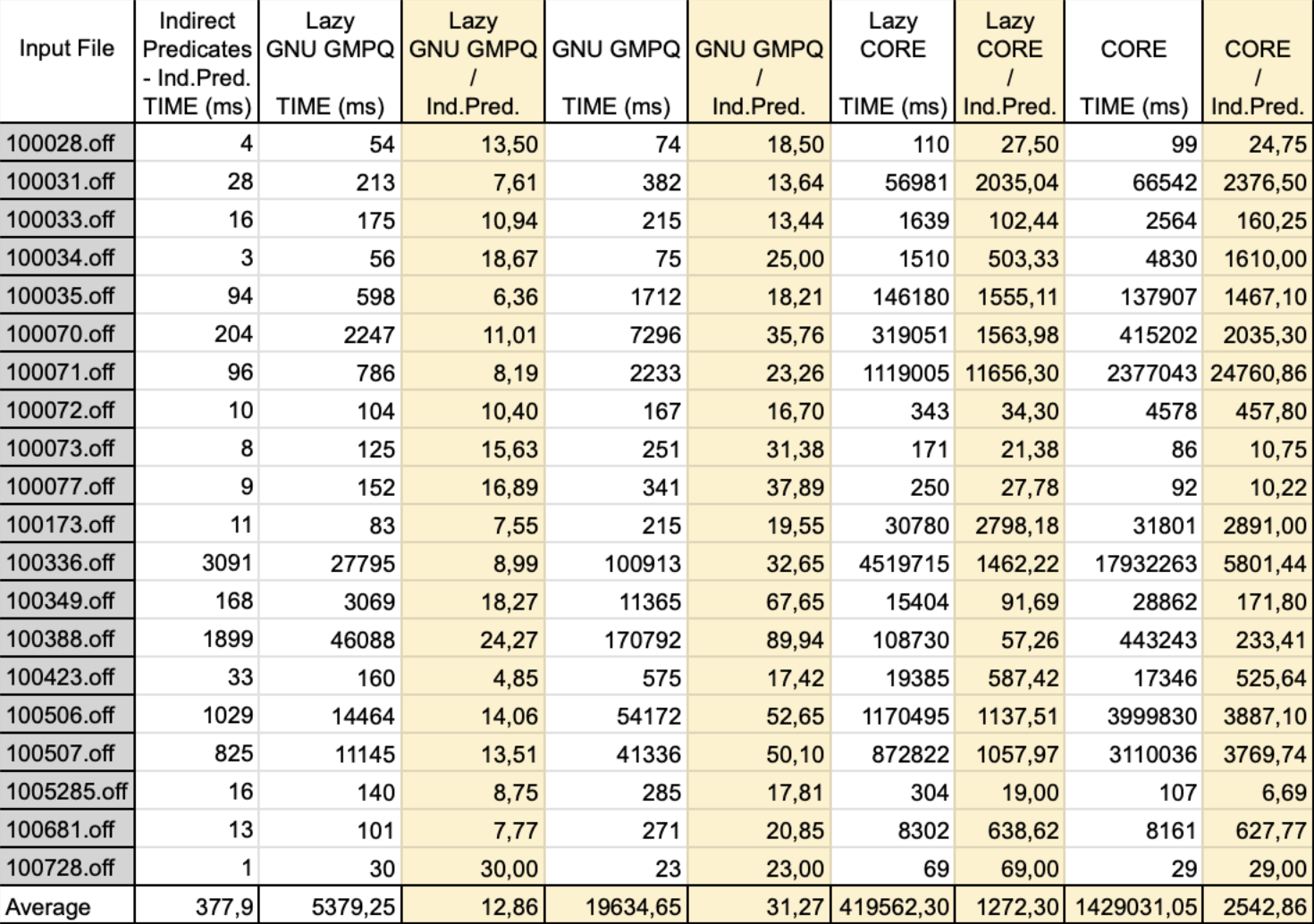}
    \caption{This table reports processing times for the first 20 models in our dataset while using our algorithm implemented with different number types as described in section \ref{sec:exact-cdt}. Yellow columns show how slower each version is when compared with the fast version based on indirect predicates.}
    \label{fig:tab_ours_gmpq_core_compare}
\end{figure}

Depending on the number type used, our tool has fairly different performances, although it remains robust in all cases.
Fig. \ref{fig:tab_ours_gmpq_core_compare} reports a comparison of the various versions for the first twenty models in our dataset. All the versions were compiled with full optimization and run on the same machine. Yellow columns show how slower each version is when compared with the fast version based on indirect predicates, revealing that CORE is thousands of times slower, even when wrapped around a lazy evaluation framework. GMP rationals perform better, but are still more than one order of magnitude slower.

\begin{figure*}
    \centering
    \includegraphics[width=\linewidth]{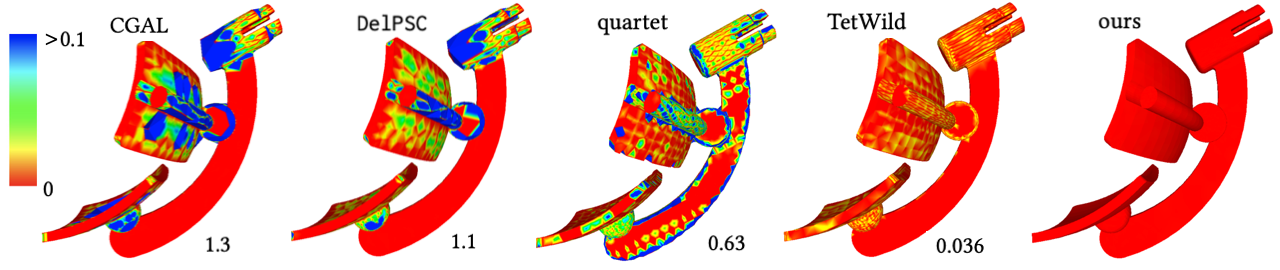}
    \caption{Tetrahedral meshes produced by different tools on the same input (Thingi10k model no. 112544). Colors depict the Hausdorff distance between the tetrahedral mesh surface and the input PLC. The maximum distance is indicated alongside each model. CGAL was configured with \emph{threshold angle}=15 and \emph{minimum radius edge ratio}=2, DelPSC with \emph{radius edge ratio}=2 and \emph{feature angle}=120, quartet with \emph{grid spacing}=0.5 and \emph{sharp features threshold angle}=170. Other methods were run with default settings. On this model tetgen fails, whereas DA2021 degenerates after rounding to floating points. Our method is the only one that \emph{exactly} conforms with the input PLC while being representable in floating point precision.}
    \label{fig:delpsc_zoom_on_approx}
\end{figure*}

\subsection{Comparison}
\label{subsec:comparison}

Our method is fundamental in all cases where downstream applications expect the mesh to be valid (i.e. no flipped or degenerate elements allowed) and exactly match the input surface. Furthermore, downstream applications hardly support exact rational coordinates, meaning that the aforementioned valid mesh must be representable using floating-point coordinates. These requirements are rather common in practice, but nonetheless strongly limit the set of appropriate tools that are comparable to ours. We compare against the Delaunay mesher \texttt{DelPSC} \cite{delpsc}, \texttt{CGAL} meshing with sharp feature preservation \rev{\cite{Rineau:2007}}, the isosurface stuffing implemented in \texttt{Quartet} \rev{\cite{Bridson:2014}}, the recent \texttt{tetwild} algorithm \rev{\cite{Hu:2018}}, the state-or-the-art software \texttt{tetgen} \cite{HSi2005CDT}, and the recent fast and robust polyhedral meshing algorithm DA2021 \cite{DiazziAttene2021}. We used the implementations provided by their authors. When mandatory input parameters are required, we \rev{tuned them to} obtain the best possible result within 1 hour. 

\texttt{DelPSC}, \texttt{CGAL}, \texttt{Quartet}, and \texttt{tetwild} produce approximate boundary conformity, even when sharp feature preservation is used. Hence, the meshes produced do not exactly match the input (see Fig. \ref{fig:delpsc_zoom_on_approx}). In principle, \texttt{CGAL} can be configured to force any non-flat edge to be preserved in the output, hence leading to an exactly conformal mesh. However, in many cases, the algorithm fails to converge with this setting. For example, the result shown in Fig. \ref{fig:delpsc_zoom_on_approx} was obtained by asking \texttt{CGAL} to preserve all edges whose incident triangle normals form an angle of at least 15 degrees. Any attempt to lower this value leads to convergence failure.

\begin{figure}
    \centering
    \includegraphics[width=\columnwidth]{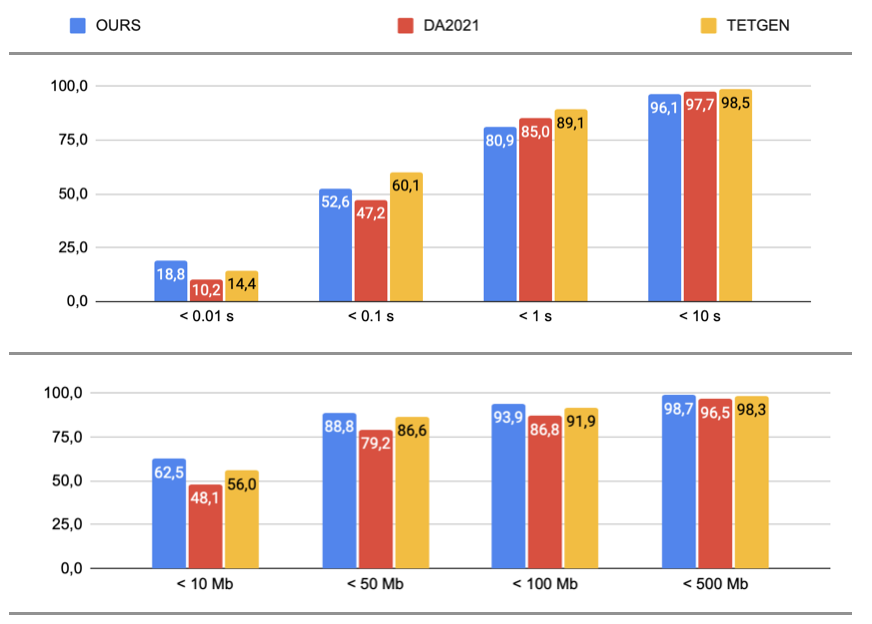}
    \caption{Time (top) and memory usage (bottom) comparisons between our method, DA2021, and \texttt{tetgen}. Each bar represents the percentage of models in the dataset that could be processed within the limit condition at its bottom.}
    \label{fig:histo_time-mem_all}
\end{figure}

We consequently focus our comparison on the remaining two algorithms, \texttt{tetgen} and DA2021, which can exactly preserve the input. On the subset of 4030 models where both these methods succeed, we compare the elapsed time and memory usage (Fig. \ref{fig:histo_time-mem_all}). In terms of performance, our algorithm is comparable with both \texttt{tetgen} and DA2021, while requiring consistently less memory. Our implementation can process 19.4\% of the models in less than 0.01 seconds, whereas \texttt{tetgen} achieves this speed for 14.4\% of the models and DA2021 for 10.2\% (file reading and writing times are excluded). In contrast, our method can process 95.2\% of the models within 10 seconds, compared to 98.5\% for \texttt{tetgen} and 97.7\% for DA2021.

The results produced by \texttt{tetgen} are always representable using floating-point coordinates, which is not surprising as \texttt{tetgen} uses this number type in its entire pipeline. Nonetheless, it fails in 378 out of 4408 models (8.6\%) in our test dataset. One might argue that \texttt{tetgen} fails because it snaps together input vertices if they are closer than a certain threshold (by default set to $10^{-8}$), introducing invalid input configurations. To counter that, we re-run the tests after having set the threshold to zero (\texttt{-T0} option in \texttt{tetgen}), which in fact increased the number of failures from 378 to 720.

Conversely, DA2021 succeeds on all the models in our dataset, but only 61.7\% of the results can be represented using floating-point coordinates with no flips and/or flattenings. When comparing against DA2021 we must first consider that this method was not designed to produce tetrahedra, though the convex polyhedral cells created can always be tetrahedrized. To do this, the authors suggest first triangulating the faces and then inserting a point at the barycenter of each non-tetrahedral cell. Approximating this barycenter using floating-point coordinates might easily invalidate the mesh. Indeed, since a polyhedral cell can be arbitrarily bad-shaped, there is no guarantee that its internal volume contains a point representable using floating-point coordinates. One might argue that the same problem may affect the Steiner points produced by our method, that is, tetrahedra may flip after having rounded the coordinates to their closest FP-representable values. While this is indeed the case, in practice our meshes are constrained Delaunay, meaning that their quality is not as arbitrarily bad as in DA2021. Thus, our meshes are still valid after rounding in 93.22\% of the cases, compared to the 61.7\% for DA2021. Furthermore, with the additional post-processing discussed in Sec. \ref{sec:fp-representation}, the mesh connectivity can be modified to eliminate most of the tets that degenerate or flip after rounding, hence increasing the percentage of valid rounded models produced by our method to 99.77\%, without any auxiliary vertices. In our implementation this latter post-processing step is optional: users can choose whether the output file must have exact rational coordinates or floating point coordinates and, in the latter case, whether to run the post-processing or not.

A theoretical question remains: is it always possible to create a valid tet mesh, possibly with Steiner points, which exactly conforms to an input surface even after rounding? The answer is no, and a counterexample is as follows.
Floating-point numbers are discrete, and points whose coordinates belong to this set form a grid in space. This grid is not regular as it is denser near the origin, but for our discussion, we may imagine it to be uniform. Let $\epsilon$ be the difference between two consecutive representable numbers, and let $p = (x,y,z)$ be one representable point. The eight points $p_i = (x + i \epsilon, y + j \epsilon, z + k \epsilon)$, with $i, j, k \in \{0,1\}$, form a cube. Now, we can take three points from its lower base (e.g. third coordinate $= z$) and other three points from the upper base (third coordinate $= z + \epsilon$) and form a Sch\"onhardt polyhedron out of them \cite{SchonPol}. This polyhedron is completely contained in the aforementioned cube and, in order to be tetrahedrized, would require at least one Steiner point within the cube itself. However, the cube does not contain any other representable point within the $\epsilon$ resolution. 

\subsection{Limitations and discussion}
\label{subsec:limitations}
While robust, our algorithm has some limitations. As one example, the use of edge-based Steiner points means we do not conform to the \emph{connectivity} of the original PLC. Even if Steiner points can be moved to the interior (Sec. 7 in \cite{HSi2011}), our tool does not support this feature as it would require an additional type of implicit point for the interior Steiner vertices. Even if our algorithm would be extended this way in the future, a PLC may still have coplanar triangular faces that form a non-Delaunay 2D triangulation and, in these cases, we cannot guarantee a conformal connectivity while still being constrained Delaunay.


Furthermore, as mentioned, the 3D Delaunay condition does not necessarily correspond to high-quality meshes \cite{alexa_harmonic} and our tool does not include a full-fledged mesh optimization phase. Mesh optimizers based on Delaunay refinement \cite{shewchuk2014higher} are mostly implemented using floating-point arithmetic that might spoil their convergence guarantees. Hence, even if our CDTs can be used to initialize these algorithms, there is no guarantee of success. To have such guarantees we would need to define new implicit points and indirect predicates to cope with mesh refinement, whose robust reimplementation is a very interesting direction for further research.

In general,  converting implicit points to floating-point coordinates with no flips remains an unsolved problem (though rare in practice), and further research is needed to discover which conditions make a PLC admit a \emph{floating-point representable} CDT.

Finally, we do not exploit modern parallel architectures. Parallelization already proved to be beneficial when computing unconstrained Delaunay tetrahedrizations \cite{marot2019one}, and we believe it represents a feasible improvement even for the constrained case.




\section{Conclusions}
We showed that, through a clever exploitation of floating point arithmetic, algorithmic efficiency and numerical robustness can be combined when calculating Steiner CDTs. Also, our research has uncovered an algorithmic issue that invalidates the theoretical guarantees of the widely used \texttt{tetgen} software. This made us able to implement a theoretically correct version of the algorithm which is also robust and fast, while bringing the failure rate from 8.6\% of the cases to zero.
This represents a significant advancement in this area because, to the best of our knowledge, no previous algorithm was capable of computing CDTs as robustly as we do.

As an interesting direction for future research, it is worth  trying to optimize our CDTs with convergence guarantees to produce high quality meshes. In principle, the algorithm in \cite{shewchuk2014higher} provides such guarantees, but in practice its floating point based implementation available in \texttt{tetgen} may easily fail or not terminate.

\begin{acks}
L. Diazzi is partly supported by the Unimore FAR Mission Oriented project 2021 ``Artificial Intelligence-based Mathematical Models and Methods for low dose CT imaging''. M. Attene is partly supported by CNR STM Project on ``Robust, Flexible and Performing Algorithms to Mesh 3D Domains''. 
This
work was partially supported by the NSF grants OAC-1835712 and CHS-1908767.
We would like to thank Silvia Sell\'{a}n and Alec Jacobson for fruitful discussions about radical number types.
\end{acks}

\bibliographystyle{ACM-Reference-Format}
\bibliography{97-bib}


\begin{thebibliography}{68}


\ifx \showCODEN    \undefined \def \showCODEN     #1{\unskip}     \fi
\ifx \showDOI      \undefined \def \showDOI       #1{#1}\fi
\ifx \showISBNx    \undefined \def \showISBNx     #1{\unskip}     \fi
\ifx \showISBNxiii \undefined \def \showISBNxiii  #1{\unskip}     \fi
\ifx \showISSN     \undefined \def \showISSN      #1{\unskip}     \fi
\ifx \showLCCN     \undefined \def \showLCCN      #1{\unskip}     \fi
\ifx \shownote     \undefined \def \shownote      #1{#1}          \fi
\ifx \showarticletitle \undefined \def \showarticletitle #1{#1}   \fi
\ifx \showURL      \undefined \def \showURL       {\relax}        \fi
\providecommand\bibfield[2]{#2}
\providecommand\bibinfo[2]{#2}
\providecommand\natexlab[1]{#1}
\providecommand\showeprint[2][]{arXiv:#2}

\bibitem[\protect\citeauthoryear{Alauzet and Marcum}{Alauzet and
  Marcum}{2014}]%
        {Alauzet:2014}
\bibfield{author}{\bibinfo{person}{Fr{\'e}d{\'e}ric Alauzet} {and}
  \bibinfo{person}{David Marcum}.} \bibinfo{year}{2014}\natexlab{}.
\newblock \showarticletitle{A closed advancing-layer method with changing
  topology mesh movement for viscous mesh generation}. In
  \bibinfo{booktitle}{\emph{Proceedings of the 22nd international meshing
  roundtable}}. Springer, \bibinfo{pages}{241--261}.
\newblock


\bibitem[\protect\citeauthoryear{Alexa}{Alexa}{2019}]%
        {alexa_harmonic}
\bibfield{author}{\bibinfo{person}{Marc Alexa}.}
  \bibinfo{year}{2019}\natexlab{}.
\newblock \showarticletitle{Harmonic Triangulations}.
\newblock \bibinfo{journal}{\emph{ACM Trans. Graph.}} \bibinfo{volume}{38},
  \bibinfo{number}{4}, Article \bibinfo{articleno}{54} (\bibinfo{date}{jul}
  \bibinfo{year}{2019}), \bibinfo{numpages}{14}~pages.
\newblock
\showISSN{0730-0301}
\urldef\tempurl%
\url{https://doi.org/10.1145/3306346.3322986}
\showDOI{\tempurl}


\bibitem[\protect\citeauthoryear{Alexa}{Alexa}{2020}]%
        {alexa_cwdt}
\bibfield{author}{\bibinfo{person}{Marc Alexa}.}
  \bibinfo{year}{2020}\natexlab{}.
\newblock \showarticletitle{Conforming Weighted Delaunay Triangulations}.
\newblock \bibinfo{journal}{\emph{ACM Trans. Graph.}} \bibinfo{volume}{39},
  \bibinfo{number}{6}, Article \bibinfo{articleno}{248} (\bibinfo{date}{nov}
  \bibinfo{year}{2020}), \bibinfo{numpages}{16}~pages.
\newblock
\showISSN{0730-0301}
\urldef\tempurl%
\url{https://doi.org/10.1145/3414685.3417776}
\showDOI{\tempurl}


\bibitem[\protect\citeauthoryear{Alliez, Cohen-Steiner, Yvinec, and
  Desbrun}{Alliez et~al\mbox{.}}{2005}]%
        {alliez:2005}
\bibfield{author}{\bibinfo{person}{Pierre Alliez}, \bibinfo{person}{David
  Cohen-Steiner}, \bibinfo{person}{Mariette Yvinec}, {and}
  \bibinfo{person}{Mathieu Desbrun}.} \bibinfo{year}{2005}\natexlab{}.
\newblock \showarticletitle{Variational tetrahedral meshing}.
\newblock In \bibinfo{booktitle}{\emph{ACM SIGGRAPH 2005 Papers}}.
  \bibinfo{pages}{617--625}.
\newblock


\bibitem[\protect\citeauthoryear{Attene}{Attene}{2019}]%
        {ipredslib}
\bibfield{author}{\bibinfo{person}{M. Attene}.}
  \bibinfo{year}{2019}\natexlab{}.
\newblock \bibinfo{title}{Indirect Predicates Library}.
\newblock
  \bibinfo{howpublished}{\url{https://github.com/MarcoAttene/Indirect\_Predicates}}.
\newblock


\bibitem[\protect\citeauthoryear{Attene}{Attene}{2020}]%
        {MAtteneIndPred}
\bibfield{author}{\bibinfo{person}{Marco Attene}.}
  \bibinfo{year}{2020}\natexlab{}.
\newblock \showarticletitle{Indirect Predicates for Geometric Constructions}.
\newblock \bibinfo{journal}{\emph{Computer-Aided Design}}
  \bibinfo{volume}{126} (\bibinfo{year}{2020}), \bibinfo{pages}{102856}.
\newblock
\urldef\tempurl%
\url{https://doi.org/10.1016/j.cad.2020.102856}
\showDOI{\tempurl}


\bibitem[\protect\citeauthoryear{Bowyer}{Bowyer}{1981}]%
        {bowyer81}
\bibfield{author}{\bibinfo{person}{Adrian Bowyer}.}
  \bibinfo{year}{1981}\natexlab{}.
\newblock \showarticletitle{Computing Dirichlet tessellations}.
\newblock \bibinfo{journal}{\emph{Comput. J.}} \bibinfo{volume}{24},
  \bibinfo{number}{2} (\bibinfo{year}{1981}), \bibinfo{pages}{162–166}.
\newblock
\urldef\tempurl%
\url{https://doi.org/10.1093/comjnl/24.2.162}
\showDOI{\tempurl}


\bibitem[\protect\citeauthoryear{Bridson and Doran}{Bridson and Doran}{2014}]%
        {Bridson:2014}
\bibfield{author}{\bibinfo{person}{Robert Bridson} {and}
  \bibinfo{person}{Crawford Doran}.} \bibinfo{year}{2014}\natexlab{}.
\newblock \showarticletitle{Quartet: A tetrahedral mesh generator that does
  isosurface stuffing with an acute tetrahedral tile}.
\newblock \bibinfo{journal}{\emph{https://github. com/crawforddoran/quartet}}
  (\bibinfo{year}{2014}).
\newblock


\bibitem[\protect\citeauthoryear{Bronson, Levine, and Whitaker}{Bronson
  et~al\mbox{.}}{2013}]%
        {Bronson:2013}
\bibfield{author}{\bibinfo{person}{Jonathan Bronson}, \bibinfo{person}{Joshua~A
  Levine}, {and} \bibinfo{person}{Ross Whitaker}.}
  \bibinfo{year}{2013}\natexlab{}.
\newblock \showarticletitle{Lattice cleaving: A multimaterial tetrahedral
  meshing algorithm with guarantees}.
\newblock \bibinfo{journal}{\emph{IEEE transactions on visualization and
  computer graphics}} \bibinfo{volume}{20}, \bibinfo{number}{2}
  (\bibinfo{year}{2013}), \bibinfo{pages}{223--237}.
\newblock


\bibitem[\protect\citeauthoryear{Burnikel, Mehlhorn, and Schirra}{Burnikel
  et~al\mbox{.}}{1996}]%
        {burnikel1996leda}
\bibfield{author}{\bibinfo{person}{C. Burnikel}, \bibinfo{person}{K. Mehlhorn},
  {and} \bibinfo{person}{S. Schirra}.} \bibinfo{year}{1996}\natexlab{}.
\newblock \bibinfo{booktitle}{\emph{The LEDA Class Real Number}}.
\newblock \bibinfo{publisher}{Max-Planck-Institut f{\"u}r Informatik}.
\newblock
\urldef\tempurl%
\url{https://books.google.it/books?id=ND5LvwEACAAJ}
\showURL{%
\tempurl}


\bibitem[\protect\citeauthoryear{Cabiddu and Attene}{Cabiddu and
  Attene}{2017}]%
        {CABIDDU2017}
\bibfield{author}{\bibinfo{person}{Daniela Cabiddu} {and}
  \bibinfo{person}{Marco Attene}.} \bibinfo{year}{2017}\natexlab{}.
\newblock \showarticletitle{epsilon-maps: Characterizing, detecting and
  thickening thin features in geometric models}.
\newblock \bibinfo{journal}{\emph{Computers \& Graphics}}  \bibinfo{volume}{66}
  (\bibinfo{year}{2017}), \bibinfo{pages}{143--153}.
\newblock
\showISSN{0097-8493}
\urldef\tempurl%
\url{https://doi.org/10.1016/j.cag.2017.05.014}
\showDOI{\tempurl}
\newblock
\shownote{Shape Modeling International 2017}.


\bibitem[\protect\citeauthoryear{Chazelle}{Chazelle}{1984}]%
        {chazelle1984}
\bibfield{author}{\bibinfo{person}{Bernard Chazelle}.}
  \bibinfo{year}{1984}\natexlab{}.
\newblock \showarticletitle{Convex partitions of polyhedra: a lower bound and
  worst-case optimal algorithm}.
\newblock \bibinfo{journal}{\emph{SIAM J. Comput.}} \bibinfo{volume}{13},
  \bibinfo{number}{3} (\bibinfo{year}{1984}), \bibinfo{pages}{488--507}.
\newblock


\bibitem[\protect\citeauthoryear{Cheng, Dey, and Levine}{Cheng
  et~al\mbox{.}}{2007}]%
        {delpsc}
\bibfield{author}{\bibinfo{person}{Siu-Wing Cheng}, \bibinfo{person}{Tamal
  Dey}, {and} \bibinfo{person}{Joshua Levine}.}
  \bibinfo{year}{2007}\natexlab{}.
\newblock \showarticletitle{A Practical Delaunay Meshing Algorithm for a Large
  Class of Domains*}.
\newblock \bibinfo{journal}{\emph{Proceedings of the 16th International Meshing
  Roundtable}}, \bibinfo{pages}{477--494}.
\newblock
\showISBNx{978-3-540-75102-1}
\urldef\tempurl%
\url{https://doi.org/10.1007/978-3-540-75103-8_27}
\showDOI{\tempurl}


\bibitem[\protect\citeauthoryear{Cheng and Dey}{Cheng and Dey}{2002}]%
        {cheng_siam}
\bibfield{author}{\bibinfo{person}{S.~W. Cheng} {and} \bibinfo{person}{T.~K.
  Dey}.} \bibinfo{year}{2002}\natexlab{}.
\newblock \showarticletitle{Quality meshing with weighted Delaunay refinement}.
\newblock \bibinfo{journal}{\emph{Proc. 13th ACM-SIAM Sympos.Discrete
  Algorithms (SODA2002)}}, \bibinfo{pages}{137--146}.
\newblock


\bibitem[\protect\citeauthoryear{Cheng, Dey, Edelsbrunner, Facello, and
  Teng}{Cheng et~al\mbox{.}}{2000}]%
        {sliver_exudation}
\bibfield{author}{\bibinfo{person}{S.~W. Cheng}, \bibinfo{person}{T.~K. Dey},
  \bibinfo{person}{H. Edelsbrunner}, \bibinfo{person}{M.~A. Facello}, {and}
  \bibinfo{person}{S.-H. Teng}.} \bibinfo{year}{2000}\natexlab{}.
\newblock \showarticletitle{Sliver exudation}.
\newblock \bibinfo{journal}{\emph{Journal of ACM}}  \bibinfo{volume}{47}
  (\bibinfo{year}{2000}), \bibinfo{pages}{883--904}.
\newblock


\bibitem[\protect\citeauthoryear{Cherchi, Livesu, Scateni, and Attene}{Cherchi
  et~al\mbox{.}}{2020}]%
        {cherchi2020}
\bibfield{author}{\bibinfo{person}{Gianmarco Cherchi}, \bibinfo{person}{Marco
  Livesu}, \bibinfo{person}{Riccardo Scateni}, {and} \bibinfo{person}{Marco
  Attene}.} \bibinfo{year}{2020}\natexlab{}.
\newblock \showarticletitle{Fast and Robust Mesh Arrangements Using
  Floating-Point Arithmetic}.
\newblock \bibinfo{journal}{\emph{ACM Trans. Graph.}} \bibinfo{volume}{39},
  \bibinfo{number}{6}, Article \bibinfo{articleno}{250} (\bibinfo{date}{nov}
  \bibinfo{year}{2020}), \bibinfo{numpages}{16}~pages.
\newblock
\showISSN{0730-0301}
\urldef\tempurl%
\url{https://doi.org/10.1145/3414685.3417818}
\showDOI{\tempurl}


\bibitem[\protect\citeauthoryear{Chew}{Chew}{1989}]%
        {chew89}
\bibfield{author}{\bibinfo{person}{L.~P. Chew}.}
  \bibinfo{year}{1989}\natexlab{}.
\newblock \showarticletitle{Constrained Delaunay triangulations}.
\newblock \bibinfo{journal}{\emph{Algorithmica}}  \bibinfo{volume}{4}
  (\bibinfo{year}{1989}), \bibinfo{pages}{97--108}.
\newblock


\bibitem[\protect\citeauthoryear{Cohen-Steiner, De~Verdiere, and
  Yvinec}{Cohen-Steiner et~al\mbox{.}}{2002}]%
        {Cohen-Steiner:2002}
\bibfield{author}{\bibinfo{person}{David Cohen-Steiner},
  \bibinfo{person}{Eric~Colin De~Verdiere}, {and} \bibinfo{person}{Mariette
  Yvinec}.} \bibinfo{year}{2002}\natexlab{}.
\newblock \showarticletitle{Conforming Delaunay triangulations in 3D}. In
  \bibinfo{booktitle}{\emph{Proceedings of the eighteenth annual symposium on
  Computational geometry}}. \bibinfo{pages}{199--208}.
\newblock


\bibitem[\protect\citeauthoryear{Cuilli{\`e}re, Francois, and
  Drouet}{Cuilli{\`e}re et~al\mbox{.}}{2013}]%
        {Cuilliere:2013}
\bibfield{author}{\bibinfo{person}{Jean-Christophe Cuilli{\`e}re},
  \bibinfo{person}{Vincent Francois}, {and} \bibinfo{person}{Jean-Marc
  Drouet}.} \bibinfo{year}{2013}\natexlab{}.
\newblock \showarticletitle{Automatic 3D mesh generation of multiple domains
  for topology optimization methods}. In \bibinfo{booktitle}{\emph{Proceedings
  of the 21st International Meshing Roundtable}}. Springer,
  \bibinfo{pages}{243--259}.
\newblock


\bibitem[\protect\citeauthoryear{Dey and Zhao}{Dey and Zhao}{2003}]%
        {dey_ma}
\bibfield{author}{\bibinfo{person}{Tamal~K. Dey} {and} \bibinfo{person}{Wulue
  Zhao}.} \bibinfo{year}{2003}\natexlab{}.
\newblock \showarticletitle{Approximating the Medial Axis from the Voronoi
  Diagram with a Convergence Guarantee}.
\newblock \bibinfo{journal}{\emph{Algorithmica}}  \bibinfo{volume}{38}
  (\bibinfo{year}{2003}), \bibinfo{pages}{179--200}.
\newblock


\bibitem[\protect\citeauthoryear{Diazzi and Attene}{Diazzi and Attene}{2021}]%
        {DiazziAttene2021}
\bibfield{author}{\bibinfo{person}{Lorenzo Diazzi} {and} \bibinfo{person}{Marco
  Attene}.} \bibinfo{year}{2021}\natexlab{}.
\newblock \showarticletitle{Convex polyhedral meshing for robust solid
  modeling}.
\newblock \bibinfo{journal}{\emph{ACM Transactions on Graphics (TOG)}}
  \bibinfo{volume}{40} (\bibinfo{year}{2021}), \bibinfo{pages}{1 -- 16}.
\newblock


\bibitem[\protect\citeauthoryear{Dobrzynski and Frey}{Dobrzynski and
  Frey}{2008}]%
        {mmg_paper}
\bibfield{author}{\bibinfo{person}{C Dobrzynski} {and} \bibinfo{person}{P.
  Frey}.} \bibinfo{year}{2008}\natexlab{}.
\newblock \showarticletitle{Anisotropic Delaunay mesh adaptation for unsteady
  simulations}. In \bibinfo{booktitle}{\emph{Proceedings of the 17th
  international Meshing Roundtable}}.
\newblock


\bibitem[\protect\citeauthoryear{Doran, Chang, and Bridson}{Doran
  et~al\mbox{.}}{2013}]%
        {Doran:2013}
\bibfield{author}{\bibinfo{person}{Crawford Doran}, \bibinfo{person}{Athena
  Chang}, {and} \bibinfo{person}{Robert Bridson}.}
  \bibinfo{year}{2013}\natexlab{}.
\newblock \showarticletitle{Isosurface stuffing improved: acute lattices and
  feature matching}.
\newblock In \bibinfo{booktitle}{\emph{ACM SIGGRAPH 2013 Talks}}.
  \bibinfo{pages}{1--1}.
\newblock


\bibitem[\protect\citeauthoryear{Du and Wang}{Du and Wang}{2003}]%
        {du:2003}
\bibfield{author}{\bibinfo{person}{Qiang Du} {and} \bibinfo{person}{Desheng
  Wang}.} \bibinfo{year}{2003}\natexlab{}.
\newblock \showarticletitle{Tetrahedral mesh generation and optimization based
  on centroidal Voronoi tessellations}.
\newblock \bibinfo{journal}{\emph{International journal for numerical methods
  in engineering}} \bibinfo{volume}{56}, \bibinfo{number}{9}
  (\bibinfo{year}{2003}), \bibinfo{pages}{1355--1373}.
\newblock


\bibitem[\protect\citeauthoryear{Du, Zhou, Carr, and Ju}{Du
  et~al\mbox{.}}{2022}]%
        {du2022}
\bibfield{author}{\bibinfo{person}{Xingyi Du}, \bibinfo{person}{Qingnan Zhou},
  \bibinfo{person}{Nathan Carr}, {and} \bibinfo{person}{Tao Ju}.}
  \bibinfo{year}{2022}\natexlab{}.
\newblock \showarticletitle{Robust Computation of Implicit Surface Networks for
  Piecewise Linear Functions}.
\newblock \bibinfo{journal}{\emph{ACM Trans. Graph.}} \bibinfo{volume}{41},
  \bibinfo{number}{4}, Article \bibinfo{articleno}{41} (\bibinfo{date}{jul}
  \bibinfo{year}{2022}), \bibinfo{numpages}{16}~pages.
\newblock
\showISSN{0730-0301}
\urldef\tempurl%
\url{https://doi.org/10.1145/3528223.3530176}
\showDOI{\tempurl}


\bibitem[\protect\citeauthoryear{Edelsbrunner and M{\"u}cke}{Edelsbrunner and
  M{\"u}cke}{1990}]%
        {symbolicperturbation}
\bibfield{author}{\bibinfo{person}{Herbert Edelsbrunner} {and}
  \bibinfo{person}{Ernst~Peter M{\"u}cke}.} \bibinfo{year}{1990}\natexlab{}.
\newblock \showarticletitle{Simulation of simplicity: a technique to cope with
  degenerate cases in geometric algorithms}.
\newblock \bibinfo{journal}{\emph{ACM Transactions on Graphics (tog)}}
  \bibinfo{volume}{9}, \bibinfo{number}{1} (\bibinfo{year}{1990}),
  \bibinfo{pages}{66--104}.
\newblock


\bibitem[\protect\citeauthoryear{Fabri and Pion}{Fabri and Pion}{2009}]%
        {fabri2009cgal}
\bibfield{author}{\bibinfo{person}{Andreas Fabri} {and}
  \bibinfo{person}{Sylvain Pion}.} \bibinfo{year}{2009}\natexlab{}.
\newblock \showarticletitle{CGAL: The computational geometry algorithms
  library}. In \bibinfo{booktitle}{\emph{Proceedings of the 17th ACM SIGSPATIAL
  international conference on advances in geographic information systems}}.
  \bibinfo{pages}{538--539}.
\newblock


\bibitem[\protect\citeauthoryear{Frey, Borouchaki, and Inria}{Frey
  et~al\mbox{.}}{1996}]%
        {Frey1996DelaunayTU}
\bibfield{author}{\bibinfo{person}{Pascal~J. Frey}, \bibinfo{person}{Houman
  Borouchaki}, {and} \bibinfo{person}{Paul Louis~George Inria}.}
  \bibinfo{year}{1996}\natexlab{}.
\newblock \showarticletitle{Delaunay Tetrahedralization using an
  Advancing-Front Approach}.
\newblock


\bibitem[\protect\citeauthoryear{George, Hecht, and Saltel}{George
  et~al\mbox{.}}{1991}]%
        {george}
\bibfield{author}{\bibinfo{person}{PL. George}, \bibinfo{person}{F. Hecht},
  {and} \bibinfo{person}{E. Saltel}.} \bibinfo{year}{1991}\natexlab{}.
\newblock \showarticletitle{Automatic mesh generator with specified boundary}.
\newblock \bibinfo{journal}{\emph{Comp Methods Appl Mechanics and Engineering}}
   \bibinfo{volume}{92} (\bibinfo{year}{1991}), \bibinfo{pages}{269--288}.
\newblock


\bibitem[\protect\citeauthoryear{Granlund}{Granlund}{1996}]%
        {gnugmp}
\bibfield{author}{\bibinfo{person}{Torbj{\"o}rn Granlund}.}
  \bibinfo{year}{1996}\natexlab{}.
\newblock \showarticletitle{Gnu mp}.
\newblock \bibinfo{journal}{\emph{The GNU Multiple Precision Arithmetic
  Library}} \bibinfo{volume}{2}, \bibinfo{number}{2} (\bibinfo{year}{1996}).
\newblock


\bibitem[\protect\citeauthoryear{Guan, Song, and Gu}{Guan
  et~al\mbox{.}}{2006}]%
        {guan}
\bibfield{author}{\bibinfo{person}{Zhenqun Guan}, \bibinfo{person}{Chao Song},
  {and} \bibinfo{person}{Yuanxian Gu}.} \bibinfo{year}{2006}\natexlab{}.
\newblock \showarticletitle{The boundary recovery and sliver elimination
  algorithms of three-dimensional constrained Delaunay triangulation}.
\newblock \bibinfo{journal}{\emph{Internat. J. Numer. Methods Engrg.}}
  \bibinfo{volume}{68}, \bibinfo{number}{2} (\bibinfo{year}{2006}),
  \bibinfo{pages}{192--209}.
\newblock
\urldef\tempurl%
\url{https://doi.org/10.1002/nme.1707}
\showDOI{\tempurl}
\showeprint{https://onlinelibrary.wiley.com/doi/pdf/10.1002/nme.1707}


\bibitem[\protect\citeauthoryear{Haimes}{Haimes}{2015}]%
        {Haimes:2015}
\bibfield{author}{\bibinfo{person}{Robert Haimes}.}
  \bibinfo{year}{2015}\natexlab{}.
\newblock \showarticletitle{MOSS: multiple orthogonal strand system}.
\newblock \bibinfo{journal}{\emph{Engineering with Computers}}
  \bibinfo{volume}{31} (\bibinfo{year}{2015}), \bibinfo{pages}{453--463}.
\newblock


\bibitem[\protect\citeauthoryear{Hu, Schneider, Wang, Zorin, and Panozzo}{Hu
  et~al\mbox{.}}{2020}]%
        {Hu:2020}
\bibfield{author}{\bibinfo{person}{Yixin Hu}, \bibinfo{person}{Teseo
  Schneider}, \bibinfo{person}{Bolun Wang}, \bibinfo{person}{Denis Zorin},
  {and} \bibinfo{person}{Daniele Panozzo}.} \bibinfo{year}{2020}\natexlab{}.
\newblock \showarticletitle{Fast Tetrahedral Meshing in the Wild}.
\newblock \bibinfo{journal}{\emph{ACM Trans. Graph.}} \bibinfo{volume}{39},
  \bibinfo{number}{4}, Article \bibinfo{articleno}{117} (\bibinfo{date}{aug}
  \bibinfo{year}{2020}), \bibinfo{numpages}{18}~pages.
\newblock
\showISSN{0730-0301}
\urldef\tempurl%
\url{https://doi.org/10.1145/3386569.3392385}
\showDOI{\tempurl}


\bibitem[\protect\citeauthoryear{Hu, Zhou, Gao, Jacobson, Zorin, and
  Panozzo}{Hu et~al\mbox{.}}{2018}]%
        {Hu:2018}
\bibfield{author}{\bibinfo{person}{Yixin Hu}, \bibinfo{person}{Qingnan Zhou},
  \bibinfo{person}{Xifeng Gao}, \bibinfo{person}{Alec Jacobson},
  \bibinfo{person}{Denis Zorin}, {and} \bibinfo{person}{Daniele Panozzo}.}
  \bibinfo{year}{2018}\natexlab{}.
\newblock \showarticletitle{Tetrahedral Meshing in the Wild}.
\newblock \bibinfo{journal}{\emph{ACM Trans. Graph.}} \bibinfo{volume}{37},
  \bibinfo{number}{4}, Article \bibinfo{articleno}{60} (\bibinfo{date}{July}
  \bibinfo{year}{2018}), \bibinfo{numpages}{14}~pages.
\newblock
\showISSN{0730-0301}
\urldef\tempurl%
\url{https://doi.org/10.1145/3197517.3201353}
\showDOI{\tempurl}


\bibitem[\protect\citeauthoryear{Jacobson, Kavan, and Sorkine-Hornung}{Jacobson
  et~al\mbox{.}}{2013}]%
        {jacobson2013robust}
\bibfield{author}{\bibinfo{person}{Alec Jacobson}, \bibinfo{person}{Ladislav
  Kavan}, {and} \bibinfo{person}{Olga Sorkine-Hornung}.}
  \bibinfo{year}{2013}\natexlab{}.
\newblock \showarticletitle{Robust inside-outside segmentation using
  generalized winding numbers}.
\newblock \bibinfo{journal}{\emph{ACM Transactions on Graphics (TOG)}}
  \bibinfo{volume}{32}, \bibinfo{number}{4} (\bibinfo{year}{2013}),
  \bibinfo{pages}{1--12}.
\newblock


\bibitem[\protect\citeauthoryear{Jamin, Alliez, Yvinec, and Boissonnat}{Jamin
  et~al\mbox{.}}{2015}]%
        {Jamin:2015}
\bibfield{author}{\bibinfo{person}{Cl{\'e}ment Jamin}, \bibinfo{person}{Pierre
  Alliez}, \bibinfo{person}{Mariette Yvinec}, {and}
  \bibinfo{person}{Jean-Daniel Boissonnat}.} \bibinfo{year}{2015}\natexlab{}.
\newblock \showarticletitle{CGALmesh: a generic framework for delaunay mesh
  generation}.
\newblock \bibinfo{journal}{\emph{ACM Transactions on Mathematical Software
  (TOMS)}} \bibinfo{volume}{41}, \bibinfo{number}{4} (\bibinfo{year}{2015}),
  \bibinfo{pages}{1--24}.
\newblock


\bibitem[\protect\citeauthoryear{Joe}{Joe}{1991}]%
        {geompack}
\bibfield{author}{\bibinfo{person}{B. Joe}.} \bibinfo{year}{1991}\natexlab{}.
\newblock \showarticletitle{GEOMPACK — A software package for the generation
  of meshes using geometric algorithms}.
\newblock \bibinfo{journal}{\emph{Adv. Engin. Software}}  \bibinfo{volume}{51}
  (\bibinfo{year}{1991}), \bibinfo{pages}{325--331}.
\newblock


\bibitem[\protect\citeauthoryear{Karamcheti, Li, Pechtchanski, and
  Yap}{Karamcheti et~al\mbox{.}}{1999}]%
        {core1999}
\bibfield{author}{\bibinfo{person}{V. Karamcheti}, \bibinfo{person}{C. Li},
  \bibinfo{person}{I. Pechtchanski}, {and} \bibinfo{person}{C. Yap}.}
  \bibinfo{year}{1999}\natexlab{}.
\newblock \showarticletitle{A Core Library for robust numeric and geometric
  computation}. In \bibinfo{booktitle}{\emph{Procs 15th ACM Symp on
  Computational Geometry (SoCG)}}. \bibinfo{pages}{351--359}.
\newblock


\bibitem[\protect\citeauthoryear{Labelle and Shewchuk}{Labelle and
  Shewchuk}{2007}]%
        {isostuff}
\bibfield{author}{\bibinfo{person}{Fran\c{c}ois Labelle} {and}
  \bibinfo{person}{Jonathan~Richard Shewchuk}.}
  \bibinfo{year}{2007}\natexlab{}.
\newblock \showarticletitle{Isosurface Stuffing: Fast Tetrahedral Meshes with
  Good Dihedral Angles}.
\newblock \bibinfo{journal}{\emph{ACM Trans. Graph.}} \bibinfo{volume}{26},
  \bibinfo{number}{3} (\bibinfo{date}{jul} \bibinfo{year}{2007}),
  \bibinfo{pages}{57–es}.
\newblock
\showISSN{0730-0301}
\urldef\tempurl%
\url{https://doi.org/10.1145/1276377.1276448}
\showDOI{\tempurl}


\bibitem[\protect\citeauthoryear{Lagae and Dutre}{Lagae and Dutre}{2008}]%
        {cdt4raytracing}
\bibfield{author}{\bibinfo{person}{A. Lagae} {and} \bibinfo{person}{P. Dutre}.}
  \bibinfo{year}{2008}\natexlab{}.
\newblock \showarticletitle{Accelerating Ray Tracing using Constrained
  Tetrahedralizations}.
\newblock \bibinfo{journal}{\emph{Computer Graphics Forum}}
  \bibinfo{volume}{4} (\bibinfo{year}{2008}), \bibinfo{pages}{1303--1312}.
\newblock


\bibitem[\protect\citeauthoryear{Lee and Lin}{Lee and Lin}{1986}]%
        {lee86}
\bibfield{author}{\bibinfo{person}{Der-Tsai Lee} {and}
  \bibinfo{person}{Arthur~K. Lin}.} \bibinfo{year}{1986}\natexlab{}.
\newblock \showarticletitle{Generalized {D}elaunay Triangulations for Planar
  Graphs}.
\newblock \bibinfo{journal}{\emph{Discrete \& Computational Geometry}}
  \bibinfo{volume}{1} (\bibinfo{year}{1986}), \bibinfo{pages}{201--217}.
\newblock


\bibitem[\protect\citeauthoryear{Li, Pion, and Yap}{Li et~al\mbox{.}}{2005}]%
        {li2005}
\bibfield{author}{\bibinfo{person}{C. Li}, \bibinfo{person}{S. Pion}, {and}
  \bibinfo{person}{C.K. Yap}.} \bibinfo{year}{2005}\natexlab{}.
\newblock \showarticletitle{Recent progress in exact geometric computation}.
\newblock \bibinfo{journal}{\emph{The Journal of Logic and Algebraic
  Programming}} \bibinfo{volume}{64}, \bibinfo{number}{1}
  (\bibinfo{year}{2005}), \bibinfo{pages}{85 -- 111}.
\newblock
\showISSN{1567-8326}
\urldef\tempurl%
\url{https://doi.org/10.1016/j.jlap.2004.07.006}
\showDOI{\tempurl}
\newblock
\shownote{Practical development of exact real number computation}.


\bibitem[\protect\citeauthoryear{Mandad, Cohen-Steiner, and Alliez}{Mandad
  et~al\mbox{.}}{2015}]%
        {Mandad:2015}
\bibfield{author}{\bibinfo{person}{Manish Mandad}, \bibinfo{person}{David
  Cohen-Steiner}, {and} \bibinfo{person}{Pierre Alliez}.}
  \bibinfo{year}{2015}\natexlab{}.
\newblock \showarticletitle{Isotopic approximation within a tolerance volume}.
\newblock \bibinfo{journal}{\emph{ACM Transactions on Graphics (TOG)}}
  \bibinfo{volume}{34}, \bibinfo{number}{4} (\bibinfo{year}{2015}),
  \bibinfo{pages}{1--12}.
\newblock


\bibitem[\protect\citeauthoryear{Marot, Pellerin, and Remacle}{Marot
  et~al\mbox{.}}{2019}]%
        {marot2019one}
\bibfield{author}{\bibinfo{person}{C{\'e}lestin Marot}, \bibinfo{person}{Jeanne
  Pellerin}, {and} \bibinfo{person}{Jean-Fran{\c{c}}ois Remacle}.}
  \bibinfo{year}{2019}\natexlab{}.
\newblock \showarticletitle{One machine, one minute, three billion tetrahedra}.
\newblock \bibinfo{journal}{\emph{Internat. J. Numer. Methods Engrg.}}
  \bibinfo{volume}{117}, \bibinfo{number}{9} (\bibinfo{year}{2019}),
  \bibinfo{pages}{967--990}.
\newblock


\bibitem[\protect\citeauthoryear{Miller, Talmor, Teng, Walkington, and
  Wang}{Miller et~al\mbox{.}}{1996}]%
        {miller1996control}
\bibfield{author}{\bibinfo{person}{Gary~L Miller}, \bibinfo{person}{Dafna
  Talmor}, \bibinfo{person}{Shang-Hua Teng}, \bibinfo{person}{Noel Walkington},
  {and} \bibinfo{person}{Han Wang}.} \bibinfo{year}{1996}\natexlab{}.
\newblock \showarticletitle{Control volume meshes using sphere packing:
  Generation, refinement and coarsening}.
\newblock \bibinfo{journal}{\emph{Proc. of 5th Intl. Meshing Roundtable}}
  (\bibinfo{year}{1996}).
\newblock


\bibitem[\protect\citeauthoryear{MMG3D}{MMG3D}{2004}]%
        {mmglib}
\bibfield{author}{\bibinfo{person}{MMG3D}.} \bibinfo{year}{2004}\natexlab{}.
\newblock \bibinfo{title}{Mmg Platform - Robust open-source and
  multidisciplinary software for remeshing}.
\newblock \bibinfo{howpublished}{\url{https://www.mmgtools.org/}}.
\newblock


\bibitem[\protect\citeauthoryear{Molino, Bridson, and Fedkiw}{Molino
  et~al\mbox{.}}{2003}]%
        {Molino:2003}
\bibfield{author}{\bibinfo{person}{Neil Molino}, \bibinfo{person}{Robert
  Bridson}, {and} \bibinfo{person}{Ronald Fedkiw}.}
  \bibinfo{year}{2003}\natexlab{}.
\newblock \showarticletitle{Tetrahedral mesh generation for deformable bodies}.
  In \bibinfo{booktitle}{\emph{Proc. Symposium on Computer Animation}},
  Vol.~\bibinfo{volume}{8}.
\newblock


\bibitem[\protect\citeauthoryear{Murphy, Mount, and Gable}{Murphy
  et~al\mbox{.}}{2001}]%
        {murphy2001point}
\bibfield{author}{\bibinfo{person}{Michael Murphy}, \bibinfo{person}{David~M
  Mount}, {and} \bibinfo{person}{Carl~W Gable}.}
  \bibinfo{year}{2001}\natexlab{}.
\newblock \showarticletitle{A point-placement strategy for conforming Delaunay
  tetrahedralization}.
\newblock \bibinfo{journal}{\emph{International Journal of Computational
  Geometry \& Applications}} \bibinfo{volume}{11}, \bibinfo{number}{06}
  (\bibinfo{year}{2001}), \bibinfo{pages}{669--682}.
\newblock


\bibitem[\protect\citeauthoryear{Rineau and Yvinec}{Rineau and Yvinec}{2007}]%
        {Rineau:2007}
\bibfield{author}{\bibinfo{person}{Laurent Rineau} {and}
  \bibinfo{person}{Mariette Yvinec}.} \bibinfo{year}{2007}\natexlab{}.
\newblock \showarticletitle{A generic software design for Delaunay refinement
  meshing}.
\newblock \bibinfo{journal}{\emph{Computational Geometry}}
  \bibinfo{volume}{38}, \bibinfo{number}{1-2} (\bibinfo{year}{2007}),
  \bibinfo{pages}{100--110}.
\newblock


\bibitem[\protect\citeauthoryear{Ruppert}{Ruppert}{1995}]%
        {ruppert:1995}
\bibfield{author}{\bibinfo{person}{J. Ruppert}.}
  \bibinfo{year}{1995}\natexlab{}.
\newblock \showarticletitle{A Delaunay Refinement Algorithm for Quality
  2-Dimensional Mesh Generation}.
\newblock \bibinfo{journal}{\emph{Journal of Algorithms}} \bibinfo{volume}{18},
  \bibinfo{number}{3} (\bibinfo{year}{1995}), \bibinfo{pages}{548--585}.
\newblock
\showISSN{0196-6774}
\urldef\tempurl%
\url{https://doi.org/10.1006/jagm.1995.1021}
\showDOI{\tempurl}


\bibitem[\protect\citeauthoryear{Ruppert and Seidel}{Ruppert and
  Seidel}{1992}]%
        {RupSei92}
\bibfield{author}{\bibinfo{person}{Jim Ruppert} {and} \bibinfo{person}{Raimund
  Seidel}.} \bibinfo{year}{1992}\natexlab{}.
\newblock \showarticletitle{On the difficulty of triangulating
  three-dimensional Nonconvex Polyhedra}.
\newblock \bibinfo{journal}{\emph{Discrete \& Computational Geometry}}
  \bibinfo{volume}{7}, \bibinfo{number}{3} (\bibinfo{date}{mar}
  \bibinfo{year}{1992}).
\newblock
\showISSN{1432-0444}
\urldef\tempurl%
\url{https://doi.org/10.1007/BF02187840}
\showDOI{\tempurl}


\bibitem[\protect\citeauthoryear{Sch\"onhardt}{Sch\"onhardt}{1928}]%
        {SchonPol}
\bibfield{author}{\bibinfo{person}{E. Sch\"onhardt}.}
  \bibinfo{year}{1928}\natexlab{}.
\newblock \showarticletitle{\"Uber die Zerlegung von Dreieckspolyedern in
  Tetraeder}.
\newblock \bibinfo{journal}{\emph{Math. Ann.}}  \bibinfo{volume}{86}
  (\bibinfo{year}{1928}), \bibinfo{pages}{309--312}.
\newblock
\urldef\tempurl%
\url{https://doi.org/10.1007/BF01451597}
\showDOI{\tempurl}


\bibitem[\protect\citeauthoryear{Shen, O'Brien, and Shewchuk}{Shen
  et~al\mbox{.}}{2004}]%
        {Shen:2004}
\bibfield{author}{\bibinfo{person}{Chen Shen}, \bibinfo{person}{James~F
  O'Brien}, {and} \bibinfo{person}{Jonathan~R Shewchuk}.}
  \bibinfo{year}{2004}\natexlab{}.
\newblock \showarticletitle{Interpolating and approximating implicit surfaces
  from polygon soup}.
\newblock In \bibinfo{booktitle}{\emph{ACM SIGGRAPH 2004 Papers}}.
  \bibinfo{pages}{896--904}.
\newblock


\bibitem[\protect\citeauthoryear{Shewchuk}{Shewchuk}{1998}]%
        {Shewchuk1998EThm}
\bibfield{author}{\bibinfo{person}{Jonathan Shewchuk}.}
  \bibinfo{year}{1998}\natexlab{}.
\newblock \showarticletitle{A Condition Guaranteeing the Existence of
  Higher-Dimensional Constrained Delaunay Triangulations}.
\newblock \bibinfo{journal}{\emph{14th Ann. ACM Symp. Comp. Geom.}}
  (\bibinfo{date}{06} \bibinfo{year}{1998}).
\newblock
\urldef\tempurl%
\url{https://doi.org/10.1145/276884.276893}
\showDOI{\tempurl}


\bibitem[\protect\citeauthoryear{Shewchuk}{Shewchuk}{1997}]%
        {shew97_pred}
\bibfield{author}{\bibinfo{person}{Jonathan~Richard Shewchuk}.}
  \bibinfo{year}{1997}\natexlab{}.
\newblock \showarticletitle{Adaptive Precision Floating-Point Arithmetic and
  Fast Robust Geometric Predicates}.
\newblock \bibinfo{journal}{\emph{Discrete \& Computational Geometry}}
  \bibinfo{volume}{18} (\bibinfo{year}{1997}), \bibinfo{pages}{305--363}.
\newblock
\urldef\tempurl%
\url{https://doi.org/10.1007/PL00009321}
\showDOI{\tempurl}


\bibitem[\protect\citeauthoryear{Shewchuk}{Shewchuk}{2000a}]%
        {shewchuk2000mesh}
\bibfield{author}{\bibinfo{person}{Jonathan~Richard Shewchuk}.}
  \bibinfo{year}{2000}\natexlab{a}.
\newblock \showarticletitle{Mesh generation for domains with small angles}. In
  \bibinfo{booktitle}{\emph{Proceedings of the sixteenth annual Symposium on
  Computational Geometry}}. \bibinfo{pages}{1--10}.
\newblock


\bibitem[\protect\citeauthoryear{Shewchuk}{Shewchuk}{2000b}]%
        {Shew00GW}
\bibfield{author}{\bibinfo{person}{Jonathan~Richard Shewchuk}.}
  \bibinfo{year}{2000}\natexlab{b}.
\newblock \showarticletitle{Sweep Algorithms for Constructing
  Higher-Dimensional Constrained Delaunay Triangulations}. In
  \bibinfo{booktitle}{\emph{Proceedings of the Sixteenth Annual Symposium on
  Computational Geometry}} (Clear Water Bay, Kowloon, Hong Kong).
  \bibinfo{publisher}{Association for Computing Machinery},
  \bibinfo{address}{New York, NY, USA}, \bibinfo{pages}{350–359}.
\newblock
\showISBNx{1581132247}
\urldef\tempurl%
\url{https://doi.org/10.1145/336154.336222}
\showDOI{\tempurl}


\bibitem[\protect\citeauthoryear{Shewchuk}{Shewchuk}{2002}]%
        {Shewchuk2002CDT}
\bibfield{author}{\bibinfo{person}{Jonathan~Richard Shewchuk}.}
  \bibinfo{year}{2002}\natexlab{}.
\newblock \showarticletitle{Constrained Delaunay Tetrahedralizations and
  Provably Good Boundary Recovery}. In \bibinfo{booktitle}{\emph{International
  Meshing Roundtable Conference}}.
\newblock


\bibitem[\protect\citeauthoryear{Shewchuk}{Shewchuk}{2003}]%
        {shewchuk2003updating}
\bibfield{author}{\bibinfo{person}{Jonathan~Richard Shewchuk}.}
  \bibinfo{year}{2003}\natexlab{}.
\newblock \showarticletitle{Updating and constructing constrained Delaunay and
  constrained regular triangulations by flips}. In
  \bibinfo{booktitle}{\emph{Proceedings of the nineteenth annual symposium on
  Computational geometry}}. \bibinfo{pages}{181--190}.
\newblock


\bibitem[\protect\citeauthoryear{Shewchuk and Si}{Shewchuk and Si}{2014}]%
        {shewchuk2014higher}
\bibfield{author}{\bibinfo{person}{Jonathan~Richard Shewchuk} {and}
  \bibinfo{person}{Hang Si}.} \bibinfo{year}{2014}\natexlab{}.
\newblock \showarticletitle{Higher-quality tetrahedral mesh generation for
  domains with small angles by constrained delaunay refinement}. In
  \bibinfo{booktitle}{\emph{Proceedings of the thirtieth annual symposium on
  Computational geometry}}. \bibinfo{pages}{290--299}.
\newblock


\bibitem[\protect\citeauthoryear{Si}{Si}{2008}]%
        {si_fem}
\bibfield{author}{\bibinfo{person}{H. Si}.} \bibinfo{year}{2008}\natexlab{}.
\newblock \showarticletitle{Adaptive tetrahedral mesh generation by constrained
  Delaunay refinement}.
\newblock \bibinfo{journal}{\emph{Internat. J. Numer. Methods Engrg.}}
  \bibinfo{volume}{75}, \bibinfo{number}{7} (\bibinfo{year}{2008}),
  \bibinfo{pages}{856--880}.
\newblock
\urldef\tempurl%
\url{https://doi.org/10.1002/nme.2318}
\showDOI{\tempurl}
\showeprint{https://onlinelibrary.wiley.com/doi/pdf/10.1002/nme.2318}


\bibitem[\protect\citeauthoryear{Si and Gärtner}{Si and Gärtner}{2005}]%
        {HSi2005CDT}
\bibfield{author}{\bibinfo{person}{Hang Si} {and} \bibinfo{person}{Klaus
  Gärtner}.} \bibinfo{year}{2005}\natexlab{}.
\newblock \showarticletitle{Meshing Piecewise Linear Complexes by Constrained
  Delaunay Tetrahedralizations}.
\newblock \bibinfo{journal}{\emph{Proceedings of the 14th International Meshing
  Roundtable}}, \bibinfo{pages}{147--163}.
\newblock
\showISBNx{3-540-25137-5}
\urldef\tempurl%
\url{https://doi.org/10.1007/3-540-29090-7_9}
\showDOI{\tempurl}


\bibitem[\protect\citeauthoryear{Si and Gärtner}{Si and Gärtner}{2011}]%
        {HSi2011}
\bibfield{author}{\bibinfo{person}{Hang Si} {and} \bibinfo{person}{Klaus
  Gärtner}.} \bibinfo{year}{2011}\natexlab{}.
\newblock \showarticletitle{3D boundary recovery by constrained Delaunay
  tetrahedralization}.
\newblock \bibinfo{journal}{\emph{Int. J. Numer. Meth. Engng}}
  \bibinfo{volume}{85} (\bibinfo{year}{2011}), \bibinfo{pages}{1341--1364}.
\newblock


\bibitem[\protect\citeauthoryear{Tournois, Wormser, Alliez, and
  Desbrun}{Tournois et~al\mbox{.}}{2009}]%
        {tournois:2009}
\bibfield{author}{\bibinfo{person}{J Tournois}, \bibinfo{person}{C Wormser},
  \bibinfo{person}{P Alliez}, {and} \bibinfo{person}{M Desbrun}.}
  \bibinfo{year}{2009}\natexlab{}.
\newblock \showarticletitle{Interleaving Delaunay Refinement and Optimization}.
\newblock \bibinfo{journal}{\emph{ACM Trans. Graphics}} \bibinfo{volume}{28},
  \bibinfo{number}{3} (\bibinfo{year}{2009}).
\newblock


\bibitem[\protect\citeauthoryear{Watson}{Watson}{1981}]%
        {watson81}
\bibfield{author}{\bibinfo{person}{David~F. Watson}.}
  \bibinfo{year}{1981}\natexlab{}.
\newblock \showarticletitle{Computing the $n$-dimensional {D}elaunay
  Tessellation with Application to {V}oronoi Polytopes}.
\newblock \bibinfo{journal}{\emph{Comput. J.}} \bibinfo{volume}{24},
  \bibinfo{number}{2} (\bibinfo{year}{1981}), \bibinfo{pages}{167--172}.
\newblock


\bibitem[\protect\citeauthoryear{Weatherill and Hassan}{Weatherill and
  Hassan}{1994}]%
        {weatherill}
\bibfield{author}{\bibinfo{person}{NP. Weatherill} {and} \bibinfo{person}{O.
  Hassan}.} \bibinfo{year}{1994}\natexlab{}.
\newblock \showarticletitle{Efficient three-dimensional Delaunay triangulation
  with automatic point creation and imposed boundary constraints}.
\newblock \bibinfo{journal}{\emph{Intnl J Num Methods in Engineering}}
  \bibinfo{volume}{37} (\bibinfo{year}{1994}), \bibinfo{pages}{2005--2039}.
\newblock


\bibitem[\protect\citeauthoryear{Xiao, Chen, Zheng, Zheng, and Wang}{Xiao
  et~al\mbox{.}}{2016}]%
        {tetbool}
\bibfield{author}{\bibinfo{person}{Zhoufang Xiao}, \bibinfo{person}{Jianjun
  Chen}, \bibinfo{person}{Yao Zheng}, \bibinfo{person}{Jianjing Zheng}, {and}
  \bibinfo{person}{Desheng Wang}.} \bibinfo{year}{2016}\natexlab{}.
\newblock \showarticletitle{Booleans of triangulated solids by a boundary
  conforming tetrahedral mesh generation approach}.
\newblock \bibinfo{journal}{\emph{Computers \& Graphics}}  \bibinfo{volume}{59}
  (\bibinfo{year}{2016}), \bibinfo{pages}{13--27}.
\newblock
\showISSN{0097-8493}
\urldef\tempurl%
\url{https://doi.org/10.1016/j.cag.2016.04.004}
\showDOI{\tempurl}


\bibitem[\protect\citeauthoryear{Zhou and Jacobson}{Zhou and Jacobson}{2016}]%
        {Thingi10K}
\bibfield{author}{\bibinfo{person}{Qingnan Zhou} {and} \bibinfo{person}{Alec
  Jacobson}.} \bibinfo{year}{2016}\natexlab{}.
\newblock \showarticletitle{Thingi10K: A Dataset of 10,000 3D-Printing Models}.
\newblock \bibinfo{journal}{\emph{arXiv preprint arXiv:1605.04797}}
  (\bibinfo{year}{2016}).
\newblock


\end{thebibliography}

\appendix

\section{Floating point filters}
\label{ap:filters}
The expression for all the versions of the \texttt{orient3D} and \texttt{inSphere} predicates can be obtained by replacing plain coordinates in Eqns. \ref{eq:orient3d} and \ref{eq:insphere} respectively with the expression of LNC coordinates as shown, for example, in Eqn. \ref{eq:orient3dlnc}.
In the following, extensions in the predicate name indicate the type of argument points (e.g. in \texttt{orient3D\_LEEE} the first point is implicit whereas the other three are explicit, in \texttt{inSphere\_LLLLL} all the points are implicit). The first implicit point is denoted with $i_1$ and is equal to $t_1 p_1 + (1-t_1) q_1$, hence its $x$ coordinate is $t_1 p_{1x} + (1 - t_1) q_{1x}$.
With this notation, we can now define the filters for all the versions of the predicates used. Filter values were calculated using \cite{MAtteneIndPred}. If the absolute value of the predicate calculated using floating point arithmetic is less than the filter value $\varepsilon_{\Delta}$, then it can be re-evaluated with more precision using intervals and, if necessary, exactly through expansion arithmetic \cite{MAtteneIndPred}.

\vspace{-\baselineskip}
\begin{small}

\begin{flalign*}
&\texttt{orient3d\_LEEE}(i_1, p_2, p_3, p_4) :
&\varepsilon_{\Delta} = 1.718625242119744 \; 10^{-13} \delta_{\Delta}^6
&&\\
&\texttt{orient3d\_LLEE}(i_1, i_2, p_3, p_4) :
&\varepsilon_{\Delta} = 2.495781359357355 \; 10^{-13} \delta_{\Delta}^6
&&\\
&\texttt{orient3d\_LLLE}(i_1, i_2, i_3, p_4) :
&\varepsilon_{\Delta} = 3.836930773104546 \; 10^{-13} \delta_{\Delta}^6
&&\\
&\texttt{orient3d\_LLLL}(i_1, i_2, i_3, i_4) :
&\varepsilon_{\Delta} = 5.68434188608081 \; 10^{-13} \delta_{\Delta}^6
\end{flalign*}

\begin{flalign*}
&\texttt{inSphere\_LEEEE}(i_1, p_2, p_3, p_4, p_5) :
&\varepsilon_{\Delta} = 5.295763827462003 \; 10^{-13} \delta_{\Delta}^7
&&\\
&\texttt{inSphere\_LLEEE}(i_1, i_2, p_3, p_4, p_5) :
&\varepsilon_{\Delta} = 2.218669692410916 \; 10^{-12} \delta_{\Delta}^8
&&\\
&\texttt{inSphere\_LLLEE}(i_1, i_2, i_3, p_4, p_5) :
&\varepsilon_{\Delta} = 9.019007762844938 \; 10^{-12} \delta_{\Delta}^9
&&\\
&\texttt{inSphere\_LLLLE}(i_1, i_2, i_3, i_4, p_5) :
&\varepsilon_{\Delta} = 3.581668295282733 \; 10^{-11} \delta_{\Delta}^{10}
&&\\
&\texttt{inSphere\_LLLLL}(i_1, i_2, i_3, i_4, i_5) :
&\varepsilon_{\Delta} =  1.991793396882719 \; 10^{-10} \delta_{\Delta}^{10}
\end{flalign*}

\end{small}

The value of $\delta_{\Delta}$ for an \texttt{orient3D} version with $n$ implicit arguments is:

\vspace{-\baselineskip}
\begin{small}

\begin{flalign*}
&\delta_{\Delta} = \max \{\delta_{\Delta i}, \delta_{\Delta e}\}
&&\\
&\delta_{\Delta i} = \max_{k \in 1..n} \{|p_{kx}|, |p_{ky}|, |p_{kz}|, |q_{kx} - p_{kx}|, |q_{ky} - p_{ky}|, |q_{kz} - p_{kz}|, |t_k|\}&&\\ 
&\delta_{\Delta e} = \max_{k \in n+1..4} \{ |p_{kx}|, |p_{ky}|, |p_{kz}|\}
\end{flalign*}

\end{small}

The value of $\delta_{\Delta}$ for an \texttt{inSphere} version with $n$ implicit arguments is:

\begin{small}
\begin{flalign*}
&\delta_{\Delta} = \max \{\delta_{\Delta i}, \delta_{\Delta e},
\delta_{\Delta w}
\}
&&\\
&\delta_{\Delta i} = \max_{k \in 1..n} \{|p_{kx}|, |p_{ky}|, |p_{kz}|, |q_{kx} - p_{kx}|, |q_{ky} - p_{ky}|, |q_{kz} - p_{kz}|, |t_k|\}&&\\ 
&\delta_{\Delta e} = \max_{k \in n+1..4} \{ |p_{kx} - p_{5x}|, |p_{ky} - p_{5y}|, |p_{kz} - p_{5z}|\}&&\\
&\delta_{\Delta w} = \left\lbrace
\begin{array}{cl}
0    & \text{~if~} n = 5\\
\max \{ |p_{5x}|, |p_{5y}|, |p_{5z}|\} & \text{otherwise}
\end{array}
\right.
\end{flalign*}

\end{small}

\section{Tetrahedron validity for gift-wrapping}
\label{app:GWpred}
With reference to section \ref{sec:giftwrap}, a tetrahedron 
$T=<t_0,t_1,t_2,w>$ 
is \emph{valid} if conditions 
$i), ii)$ and 
$iii)$ hold.
Condition $i)$ is equivalent to checking whether \texttt{orient3d}($t_0,t_1,t_2,w$) > 0.
Condition $ii)$ holds if, for each triangle $\tau \in \partial C_{curr}$, one of the following holds:
\begin{itemize}
    \item $\tau$ and $T$ share three vertices (i.e. $\tau$ is a face of $T$);
    \item $\tau$ and $T$ share two vertices \textbf{and} at least one of the tets obtained by replacing one of the unshared vertices in $T$ with the unshared vertex in $\tau$ have negative volume;
    \item $\tau$ and $T$ share one vertex \textbf{and} the other two vertices of $\tau$ are not contained in the volume of $T$ \textbf{and} no edge of $\tau$ intersects a face of $T$ \textbf{and} no edge of $T$ intersects $\tau$ except for the common vertex;
    \item $\tau$ and $T$ have no common vertices \textbf{and} no vertex of $\tau$ is contained in the volume of $T$ \textbf{and} no edge of $\tau$ intersects a face of $T$ \textbf{and} no edge of $T$ intersects $\tau$.
\end{itemize}

Note that this approach requires \emph{point in tetrahedron} and \emph{segment-triangle intersection} tests only. All can be checked exactly through simple calls to \texttt{orient3d} (see Algorithms \ref{alg:pointintet} and \ref{alg:segtriint}).

\begin{algorithm}

\caption{point$\_$in$\_$tetrahedron($p$, $T$)}
\label{alg:pointintet}

\textbf{Input:}\\
$p$: query point to be checked;\\
$T = <t_0,t_1,t_2,t_3>$: reference tetrahedron.\\

\textbf{Output:}\\
\textbf{true} if $p$ belongs to the volume of $T$ (including its boundary).

\hrulefill \\
\SetAlgoLined
\vspace{0.1em}

\lIf{ \texttt{orient3d$(t_0,t_1,t_2,p) < 0$ }}{ \textbf{return false}  } 
\lIf{ \texttt{orient3d$(t_0,t_1,p,t_3) < 0$ }}{ \textbf{return false}  } 
\lIf{ \texttt{orient3d$(t_0,p,t_2,t_3) < 0$ }}{ \textbf{return false}  } 
\lIf{ \texttt{orient3d$(p,t_1,t_2,t_3) < 0$ }}{ \textbf{return false}  } 
\textbf{return true}

\end{algorithm}

\begin{algorithm}

\caption{segment$\_$intersects$\_$triangle($s$, $t$)}
\label{alg:segtriint}

\textbf{Input:}\\
$s = <s_1, s_2>$: segment;\\
$t = <v_0,v_1,v_2>$: triangle.\\

\textbf{Output:}\\
\textbf{true} if $s$ and $t$ intersect while not being coplanar.

\hrulefill \\
\SetAlgoLined
\vspace{0.1em}

\lIf{ \texttt{orient3d$(v_0,v_1,v_2,s_1)$ = orient3d$(v_0,v_1,v_2,s_2)$ }}{ \textbf{return false}  } 

\lIf{ \texttt{orient3d$(v_0,v_1,s_1,s_2)$ * orient3d$(v_1,v_2,s_1,s_2)$ < 0}}{ \textbf{return false}  }

\lIf{ \texttt{orient3d$(v_1,v_2,s_1,s_2)$ * orient3d$(v_2,v_0,s_1,s_2)$ < 0}}{ \textbf{return false}  }

\lIf{ \texttt{orient3d$(2_0,v_0,s_1,s_2)$ * orient3d$(v_0,v_1,s_1,s_2)$ < 0}}{ \textbf{return false}  }

\textbf{return true}

\end{algorithm}

Condition $iii)$ holds if, for each vertex $v$ in the half-cavity, either $v$ is not in the circumsphere of $T$ or it is not visible from within $T$. 
$v$ is not visible from within $T$ even if just a portion of $T$'s relative interior is occluded by the initial half-cavity  boundary $\partial C_i$. Hence, in order for $v$ to be visible, all the internal points in $T$ must be visible from $v$. Equivalently, $v$ is visible from within $T$ if and only if, for each triangle $\tau$ in $\partial C_i$, $I(CH(T \cup v)) \cap \tau = \emptyset$, where $I(.)$ denotes the \emph{interior} operator and $CH(.)$ denotes the \emph{convex hull} operator. Because $CH(T \cup v)$ can be made of one or two tetrahedra, the previous approach to detect triangle-tetrahedron intersection can be reused.

\end{document}